\documentclass{article}

\usepackage{arxiv}

\usepackage[utf8]{inputenc} 
\usepackage[T1]{fontenc}    
\usepackage{hyperref}       
\usepackage{url}            
\usepackage{booktabs}       
\usepackage{amsfonts}       
\usepackage{nicefrac}       
\usepackage{microtype}      
\usepackage{lipsum}		
\usepackage{cite}
\usepackage{indentfirst}
\usepackage{graphicx} 
\usepackage{float} 
\usepackage{subfigure} 
\usepackage{amsmath}
\usepackage{mathtools}

\title{Continuous-Time Systems for Solving 0-1 Integer Linear Programming Feasibility Problems}

\author{
  Chengrui~Li\\
  $^1$Wu Yuzhang Honors College\\
  Sichuan University\\
  Chengdu, Sichuan, China\\
  $^2$University of Tennessee\\
  Knoxville, TN 37996, USA \\
  \texttt{cnlichengrui@foxmail.com} \\
   \And
  Bruce J.\ MacLennan\thanks{Contact author} \\
  Department of Electrical Engineering \& Computer Science\\
  University of Tennessee\\
  Knoxville, TN 37996, USA \\
  \texttt{maclennan@utk.edu} \\
  }

\begin{document}
\maketitle

\begin{abstract}
  The 0-1 integer linear programming feasibility problem is an important NP-complete problem. This paper proposes a continuous-time dynamical system for solving that problem without getting trapped in non-solution local minima. First, the problem is transformed to an easier form in linear time. Then, we propose an ``impulse algorithm'' to escape from local traps and show its performance is better than randomization for escaping traps. Second, we present the time-to-solution distribution of the impulse algorithm and compare it with exhaustive search to see its advantages. Third, we show that the fractional size of the basin of attraction of the global minimum is significantly larger than $2^{-N}$, the corresponding discrete probability for exhaustive search. Finally, we conduct a case study to show that the location of the basin is independent of different dimensions. These findings reveal a better way to solve the 0-1 integer linear programming feasibility problem continuously and show that its cost could be less than discrete methods in average cases.
\end{abstract}


\section{Introduction}
\noindent The 0-1 integer linear programming (ILP) feasibility problem is one of a relatively large class of NP-complete problems \cite{karp1972reducibility}. Unlike the optimization version of 0-1 ILP, the feasibility version does not have an objective function that needs to be optimized and its only aim is to satisfy the given constraints. Although it might be easy to solve quickly in practical applications, because the constraints of many real world problems have some special properties (e.g. the coefficients are only 0 or 1, and 0 weights are far more frequent than 1 weights), in general the complexity of the purely mathematical problem is $2^N$.

In 1971, Cook proved the first NP-complete problem, Boolean satisfiability (SAT) \cite{cook1971complexity}. In the following year, Karp identified twenty-one NP-complete problems \cite{karp1972reducibility}, which have a kind of hierarchy; that is, all of these problems can be reduced to each other. For instance, the vertex cover problem can be reduced to the clique problem, and the latter can be reduced to SAT, where SAT is the most basic NP-complete problem up to now. The time complexity of a reduction is polynomial \cite{karp1972reducibility, ladner1975structure}, but this reduction process might give a larger sized problem (also polynomially larger). Therefore, sometimes it might be undesirable to reduce a problem to a more general NP-complete problem if we just want to adopt an exhaustive search method. This is also why many investigators are interested in finding special methods for solving, for example, Hamiltonian cycle problems (directed or undirected), even though they can be reduced to SAT. Another reason is that perhaps more practical problems can be easily described as those higher level problems, such as Hamiltonian cycle, rather than by reduction to SAT. The situation is similar for 0-1 ILP feasibility problem (e.g. the facility location problem).

The 0-1 ILP feasibility problem is positioned at the second basic level of Karp's twenty-one NP-complete problems, at the same level as the clique and 3-SAT problems \cite{karp1972reducibility}. Further, no other of these twenty-one problems can be reduced to it, but many other NP-complete problems have many reductions to them. Subsequent research has proved that many interesting problems are NP-complete, such as minesweeper \cite{kaye2000minesweeper} on the Windows operating system, and some simplified (domain restricted) NP-complete problems \cite{garey1974some}.

Generally speaking, the term ``0-1 integer linear programming'' refers to the optimization version of the problem, which is composed of an objective function ($\max$ or $\min$) and a series of linear constraints. Compared with the feasibility version, the constraints of the optimization version are relatively easier to satisfy. Thus, a very common idea, the penalty function method, is widely used in many practical problems. Because of the easily satisfied constraints, in most cases very large penalty terms are able to effectively optimize the objective function when finding the maximum or minimum. In addition, many optimization problems are not very strict. In other words, sometimes an application is able to accept a relatively good result even if it is not the best. Hence, a trade-off often happens between time-cost and the extent of the optimization. When it comes to the feasibility version, however, the constraints are more complex (not a sparse matrix), and the constraints have to be satisfied strictly. The critical requirements of such problems increase the degree of difficulty.

Because there are very few papers that use continuous-time dynamical systems to solve the 0-1 ILP feasibility problem in a general sense, this paper will introduce its continuous-time method with a strategy to decrease the time complexity below exponential and to escape local traps. The last few sections also analyze the advantages of this method and the possibility of reducing its time complexity.

\section{Related Work}
\noindent In recent years many discrete problems have been solved by continuous methods, especially decision problems, because  continuous-time dynamical systems operate similarly to our brain. The state of neurons change continuously but their outputs are usually binary --- fire or not --- which depends on the real-time difference between the present state and a threshold \cite{zaremba2014recurrent}. Most of continuous-time complex systems can be treated as neural networks, and particularly when the neurons are connected with feedback paths and the input is a time sequence, this kind of network is called a Recurrent Neural Network (RNN). RNNs have been applied in many fields, such as neuroscience \cite{barak2017recurrent} and chaotic physical system \cite{huang2017once, graves2013speech}.

\begin{figure}[htbp]
  \centering
  \subfigure[]{
    \includegraphics[width=0.45\textwidth]{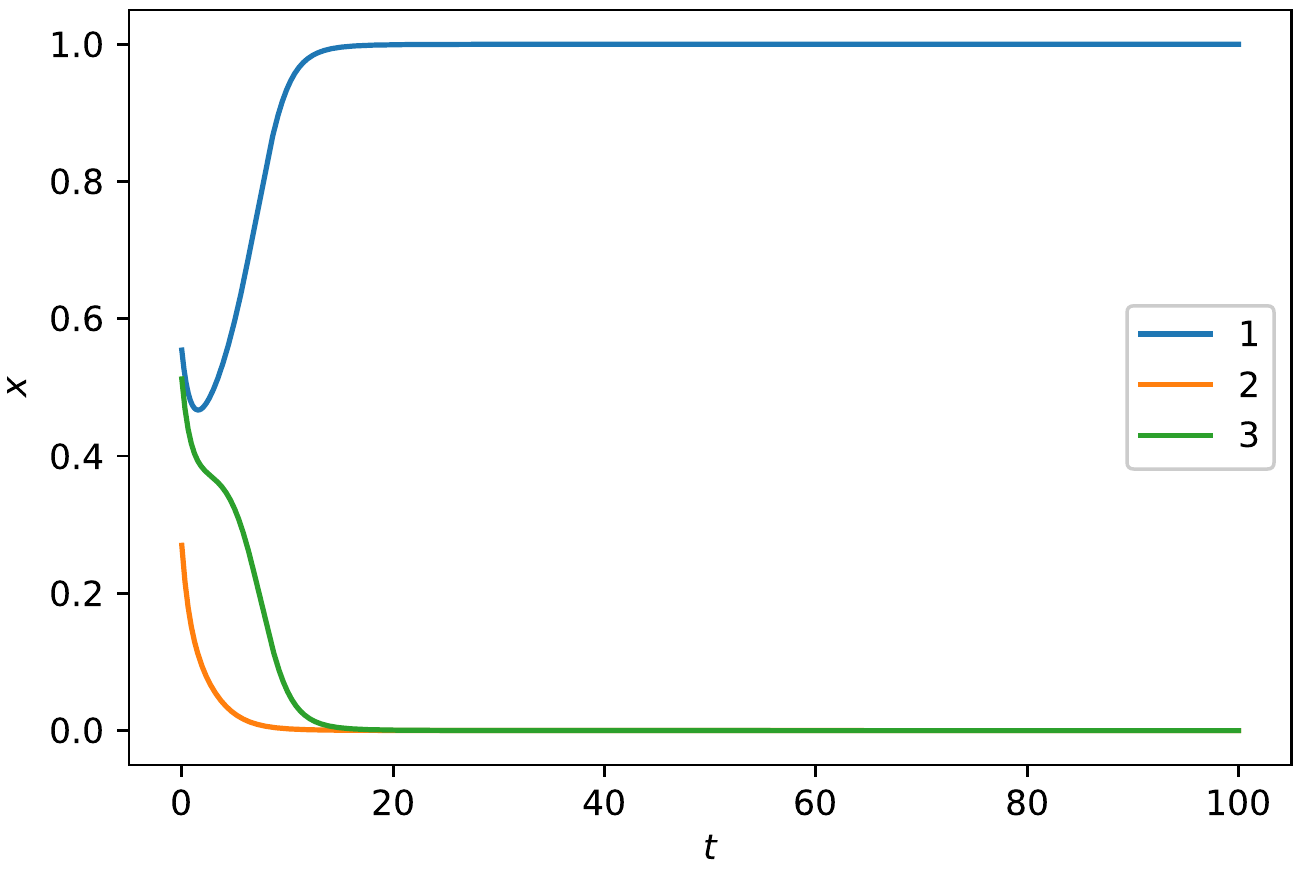}
    \label{Fig.1.sub.a}}
  \subfigure[]{
    \includegraphics[width=0.45\textwidth]{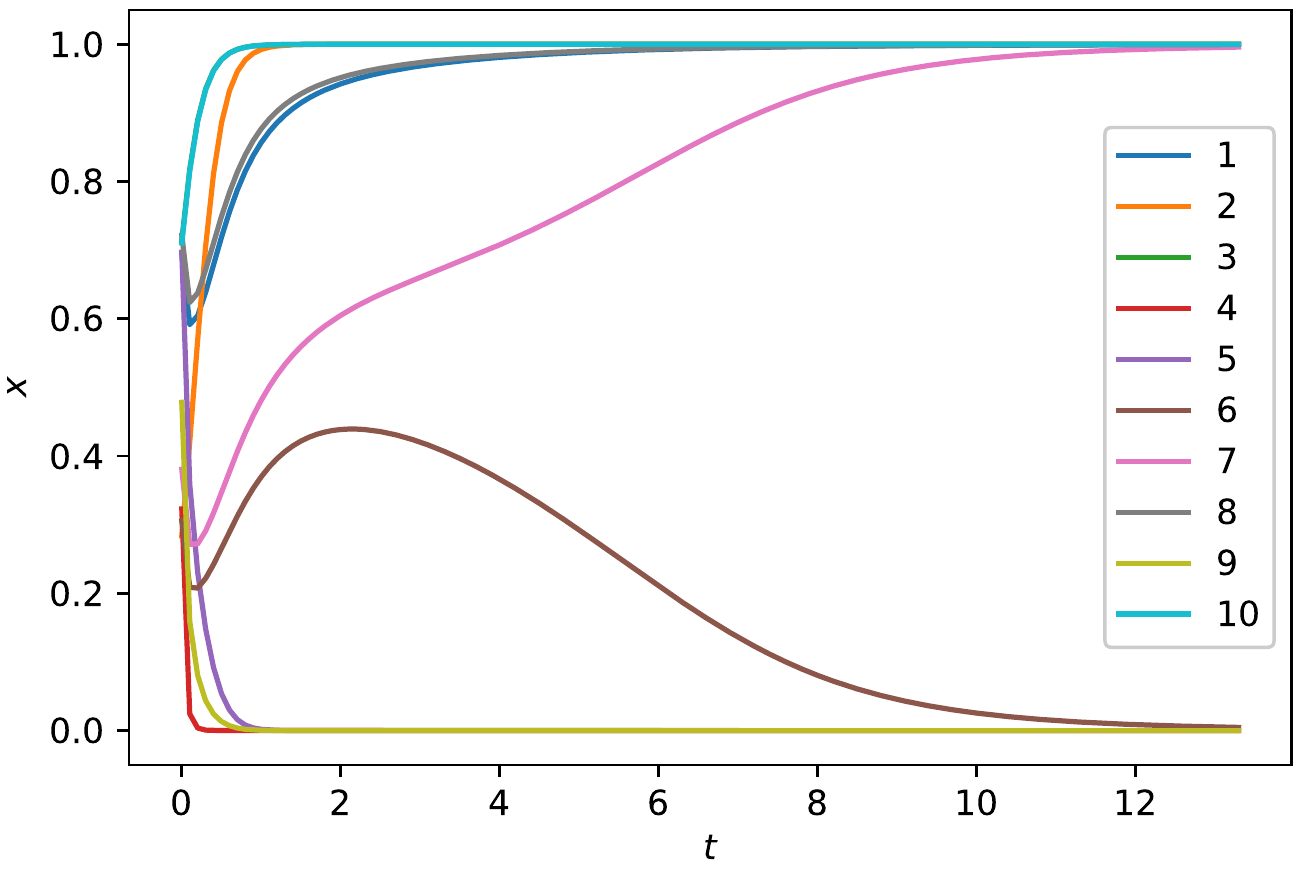}
    \label{Fig.1.sub.b}}
  \subfigure[]{

    \includegraphics[width=0.45\textwidth]{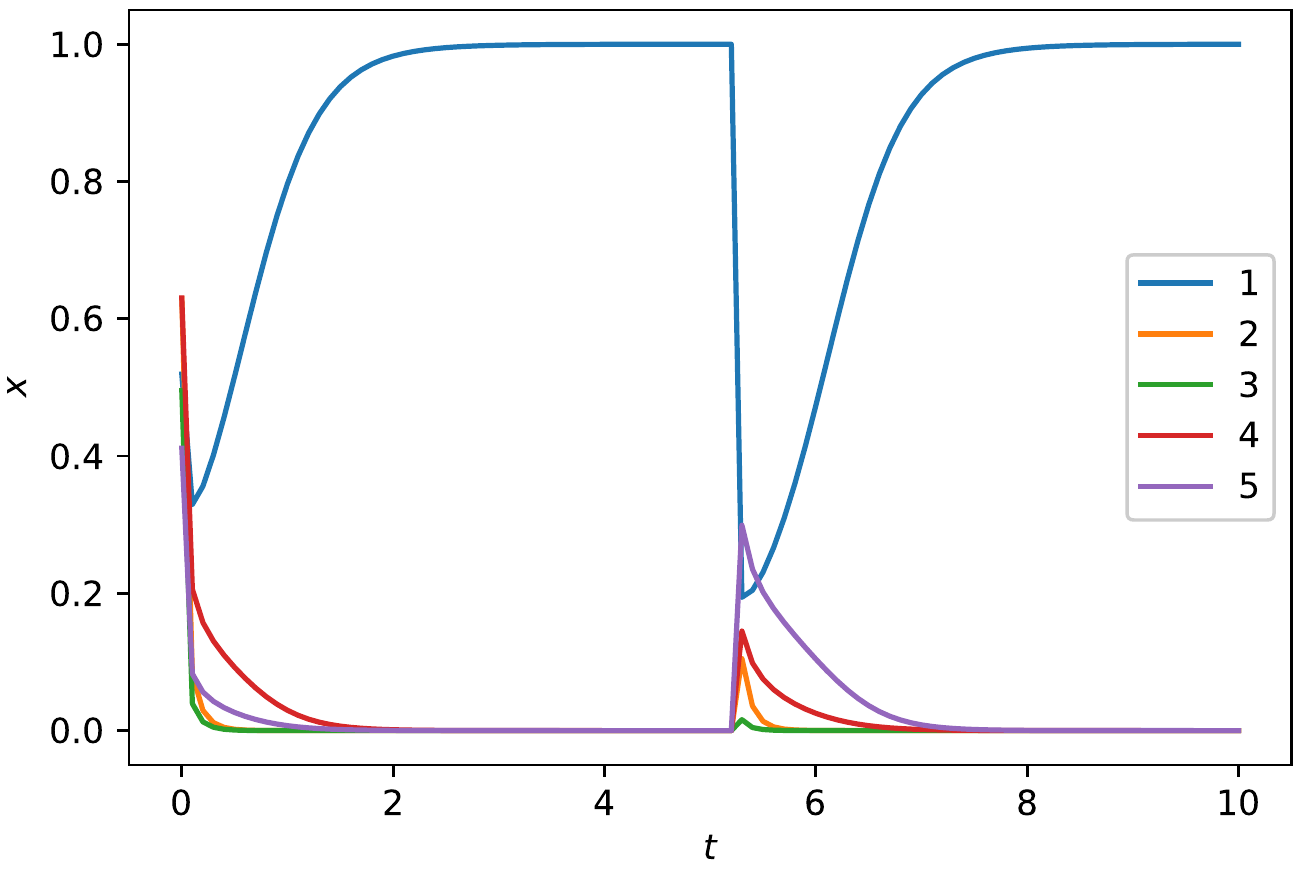}
    \label{Fig.1.sub.c}}
  \subfigure[]{
    \includegraphics[width=0.45\textwidth]{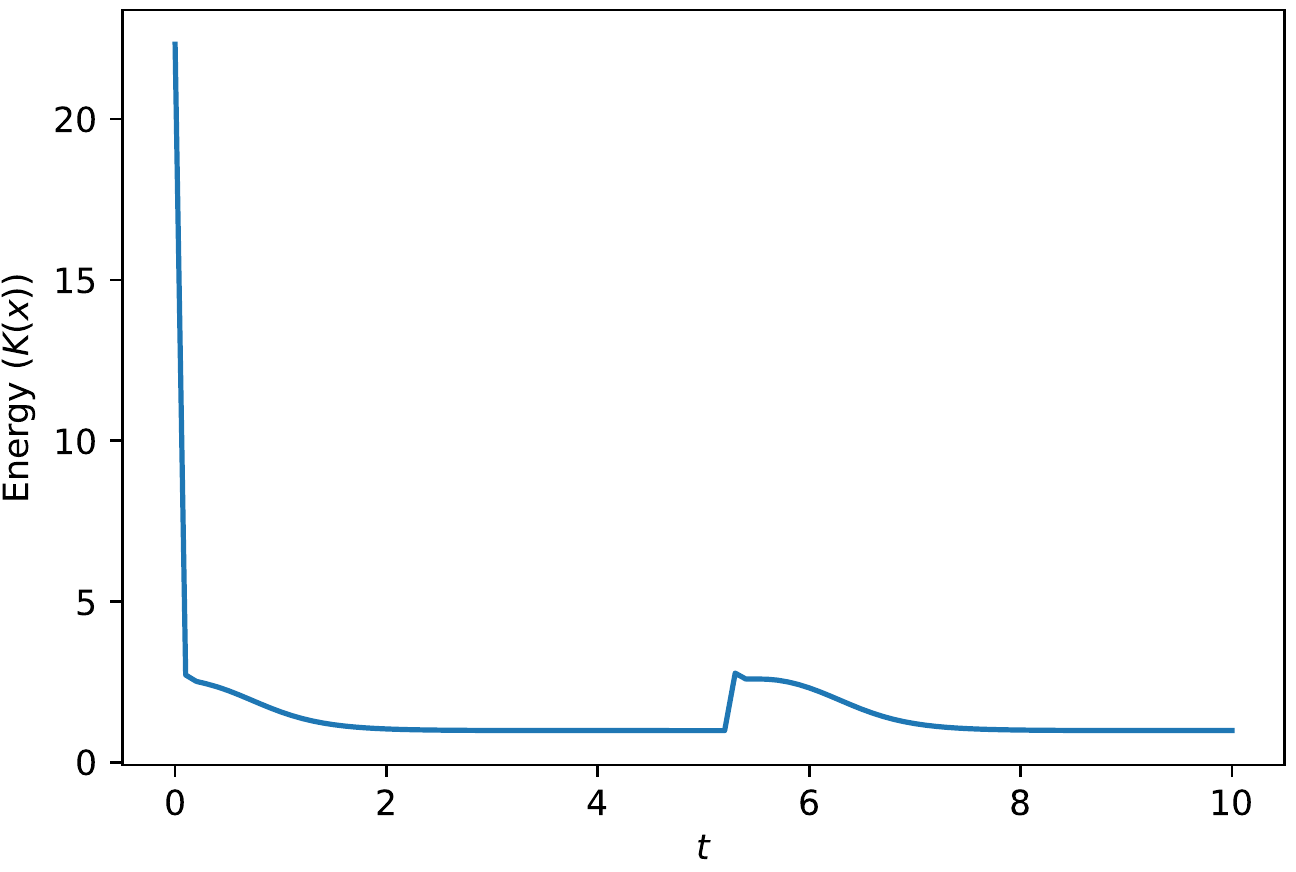}
    \label{Fig.1.sub.d}}
  \caption{Examples of Hopfield Network. (a) An example of solving a problem that has three variables, one constraint, and the coefficients are 0 or 1. The network always converges to the correct result. (b) An example of solving a problem that has ten variables, one constraint, and the coefficients could be any non-negative integer less than or equal to 3. The network sometimes converges to the result successfully. (c) An example of solving a problem that has five variables, one constraint, and the coefficients could be any non-negative integer less than or equal to 3. This time the network fails to converge to the result, in spite of the state being randomized in the local trap at $t \approx 5$. (d) The energy of the network in (c).}
  \label{Fig.1}
\end{figure}

When it comes to combinatorial problems, continuous methods have been investigated, but most of them are for solving incomplete versions of a specific problem under some special cases or for the optimization version of 0-1 ILP. For example, Tagliarini et al.\ reviewed the use of neural networks for optimization \cite{tagliarini1991optimization}. If all of the coefficients are only 0 or 1, a Hopfield neural network is able to solve the satisfiability problem continuously (e.g. Figure \ref{Fig.1}(a)). In fact, following their construction, we can also build a neural network for the general 0-1 ILP feasibility problem, but the local traps (non-solution minima) and the convergence conditions largely depend on the structure of the specific problem instance (e.g. Figure \ref{Fig.1}(b--d)). Impagliazzo et al.\ solve the 0-1 ILP feasibility problem with exponential speedup over exhaustive search in the special case where the number of constraints is a multiple of the number of variables; this is accomplished by a reduction to the vector domination problem \cite{Impagliazzo2014ILP}. Other research, including 0-1 \cite{de2012continuous} / integer \cite{genova2011linear} / mixed \cite{floudas2005mixed, di2007linear, castro2005new} programming, have usually addressed the optimization version of the problem.

If one reduces the 0-1 ILP feasibility problem to SAT, then the new problem can be solved continuously without getting caught in local traps. For example, Ercsey-Ravasz and Toroczkai added exponentially growing auxiliary terms to each constraint, so that the network is able to escape from traps and converge to the $k$-SAT solution, possibly after a period of chaotic behavior \cite{ercsey2011optimization}. Moln\'ar and Ercsey-Ravasz designed an asymmetric continuous-time neural network to solve $k$-SAT problems \cite{molnar2013asymmetric}, and this kind of network could also be used to solve the MaxSAT problem with good analog performance \cite{molnar2018continuous}. These methods inspired us to design a similar continuous-time dynamical system to solve the 0-1 ILP feasibility problem directly, since the reduction to 3-SAT would enlarge the size of the problem.

\section{General Algorithm}
\subsection{Problem transformation}
\noindent The 0-1 ILP feasibility problem can be described as: given an integer matrix $C$ and an integer vector $\boldsymbol{d}$, the goal is to find an unknown binary vector $\boldsymbol{x}$ that satisfies $C\boldsymbol{x} = \boldsymbol{d}$ \cite{karp1972reducibility}. The number of rows $M$ of $C$ represents the number of constraints, and the number of columns $N$ of $C$ represents the number of  variables. For example, if there are three constraints and five variables,
\[
\begin{cases}
    3x_1 + 10x_2 + 6x_3 + 14x_4 + 8x_5 &=\ \  17\\
    7x_1 + 4x_2 + 30x_3 + 0x_4 + x_5 &=\ \  38\\
    19x_1 + 4x_2 + 0x_3 + 5x_4 + 9x_5 &=\ \  28
\end{cases}
\]
the problem can be written as:
\begin{equation}\label{Eqn.1}
C = \left[
    \begin{array}{ccccc}
    3 & 10 &6 & 14 & 8 \\
    7 & 4 & 30 & 0 & 1 \\
    19  & 4 & 0 & 5 & 9 \\
    \end{array}
\right]
,~~~~\boldsymbol{d} = [17, 38, 28]^T
\end{equation}
whose solution could be $\boldsymbol{x^*} = [1, 0, 1, 0, 1]^T$. Here, the elements of $C$ and $\boldsymbol{d}$ are all non-negative.

To extend this method to $C \in \mathcal{M}_{M\times N}(\mathbb{Z})$, one can always substitute $(1 - y_i)$ for those $x_i$ whose corresponding coefficients are negative. This transformation gives a new matrix, and when applying the following method, these transformed variables change in the opposite direction to its corresponding constraint during the continuous updating and iteration process. More specifically, substitute $(1 - y_{mi})$ for $x_{mi}$ and keep in mind that $y_{mi}$ are used for computation and $x_i = 1 - y_i$ represent the values of the original variables, so that the $x_i$ across all constraints are identical with each other. For example, consider the following problem.
\[
\begin{cases}
    2x_1 + 8x_2 + 4_3 &=\ \  10\\
    3x_1 -2x_2 + 5x_3 &=\ \  1\\
\end{cases}
\]
In this case we replace $x_2$ in the second constraint by $y_{22} = (1 - x_2)$, and the problem is converted to
\[
\begin{cases}
    2x_1 + 8x_2 + 4_3 &=\ \  10\\
    3x_1 + 2y_{22} + 5x_3 &=\ \  3\\
\end{cases}
\]
If the incremental update for $x_2$ is $0.3$ in the first constraint and for $y_{22}$ is $0.2$ in the second constraint, then the final incremental update for $x_2$ should be the average of $0.3$ and $-0.2$. The updated value of $y_{22}$ should be $y'_{22}=(1 - x'_2)$ where $x'_2$ is the updated value of $x_2$.
This transformation costs only linear time $\mathcal{O}(N)$ but can decrease the logical complexity and increase the conciseness to a great extent, especially in a high-dimensional space. The following experiments are all conducted with $C \in \mathcal{M}_{M\times N}(\mathbb{N})$ and $\boldsymbol{d}$ for which solutions exist.

If $M = N$ and a solution exists, then Gaussian elimination is the optimal method to find it, and its complexity is $\mathcal{O}(N^3)$. On the other hand we are interested in the case $N \gg M$ ($M$ can be assumed as a constant) and whether there is an algorithm with complexity less than that of exhaustive search, $\mathcal{O}(2^N)$.
The exhaustive search method is very stable because it is not affected by the number of constraints $M$ and the range of the coefficients. Hence, exhaustive search ($\mathcal{O}(2^N)$) will be the comparison baseline in the following sections.

The satisfiability problem for linear equations $C\boldsymbol{x} = \boldsymbol{d}$ is polynomial over the reals, rationals, and integers, but NP-complete over the natural numbers \cite{bockmayr2001snc}.
In fact, the feasibility problem for integer linear programming is equivalent to the problem of solving a system linear Diophantine equations, which can be solved in polynomial time by computing either the Hermite or Smith normal forms of $C$ \cite{bockmayr2001snc}. However, if the solutions are restricted to $\{0,1\}^N$ then the problem is NP-complete, as Karp showed \cite{karp1972reducibility}.

\subsection{Dynamical system model}
\noindent The first step for finding the solution is to build an energy function $K(\boldsymbol{x})$ that  attains its global minimum only at the solution. For this purpose, two criteria should be taken into consideration: (1) the continuous solution to the equation and (2) the constraint that all the variables be binary. Thus, two terms are included:
\begin{equation}\label{Eqn.2}
  \begin{split}
  K_m(\boldsymbol{x}) &= \frac{1}{2} \left( \frac{d_m - \sum_{i=1}^{N}c_ix_i}{\sum_{i=1}^Nc_i} \right)^2 + \frac{1}{2\sum_{i=1}^Nc_i}\sum_{i=1}^Nc_i[x_i(1 - x_i)]^2\\
  K(\boldsymbol{x}) &= \frac{1}{M}\sum_{m=1}^MK_m
  \end{split}
\end{equation}
where $m$ means the $m$-th constraint, $c_i$ in $K_m$ is the coefficient of $x_i$ in the $m$-th constraint (i.e., $C_{mi}$), and $d_m$ is the $m$-th element of $\boldsymbol{d}$. The energy function is scaled by $\sum_{i=1}^Nc_i$ so that its magnitude is stable across different coefficient ranges. The first term will be $0$ for any continuous solution, and the second term will be $0$ for any binary vector $\boldsymbol{x}$. Thus, the energy function $K(\boldsymbol{x})$ attains its global minimum $0$ when $\boldsymbol{x} = \boldsymbol{x^*}$.

Gradient descent on the energy surface is then defined:
\begin{equation}\label{Eqn.3}
  \begin{split}
  &\frac{dx_j}{dt} = -\frac{\partial K(\boldsymbol{x})}{\partial x_j} = -\frac{1}{M} \sum_{m = 1}^{M} \frac{\partial K_m(\boldsymbol{x})}{\partial x_j}\\
  &\frac{\partial K_m(\boldsymbol{x})}{\partial x_j} = \frac{c_j}{\sum_{i=1}^Nc_i} \left[ \frac{d - \sum_{i=1}^{N}c_ix_i}{\sum_{i=1}^Nc_i} + x_j(1 - x_j)(2x_j - 1) \right]
  \end{split}
\end{equation}

Gradient descent is the basic iterative process to find the minimum (Equation \eqref{Eqn.3}),
but $K(\boldsymbol{x})$ is a 4-th order polynomial, so there are two minima along each dimension. Therefore, we cannot guarantee that only one minimum exists in the 0-1 hypercube $[0,1]^N$. Given this challenge, we need a mechanism embedded in the dynamics for detecting and escaping local traps (non-solution minima).

\subsection{Local trap detection mechanism}
\noindent One approach would be to follow the $k$-out-of-$n$ rules presented by Tagliarini et al. \cite{tagliarini1991optimization} to construct and to rewrite the energy function as follows:
\begin{equation}\label{Eqn.4}
\begin{split}
    K_m(\boldsymbol{s})
    & = \left[d_m - \sum_{i=1}^{N}c_ig(s_i)\right]^2 + \sum_{i=1}^{N}c_i^2g(s_i)[1 - g(s_i)]\\
    & = d_m^2 - \frac{1}{2}\sum_{i=1}^{N}\sum_{j=1, j\neq i}^{N} -2c_ic_jg(s_i)g(s_j) - \sum_{i=1}^{N}(2d_m - c_i^2)g(s_i)\\
    & \sim - \frac{1}{2}\sum_{i=1}^{N}\sum_{j=1, j\neq i}^{N} -2c_ic_jg(s_i)g(s_j) - \sum_{i=1}^{N}(2d_m - c_i^2)g(s_i)
\end{split}
\end{equation}
where they eliminated the constant $d_m^2$ to fit the form of the energy function for Hopfield's continuous model \cite{tagliarini1991optimization}:
\begin{equation}\label{Eqn.5}
\frac{ds_j}{dt} = -\frac{s_j}{\eta_i} + \sum_{i=1}^{N}T_{ij}g(s_i) + I_j
\end{equation}
where $x_i = g(s_i) = \frac{1}{1 + e^{-s_i}}$, $T_{ij} = \begin{cases}-2c_ic_j& i \neq j\\0& i = j\end{cases}$, and $I_j = 2d_m - c_j^2$.

We, however, retained the constant for detecting the global minimum, because the global minimum of our $K(\boldsymbol{x})$ (Equation \eqref{Eqn.2}) is $0$ under all conditions. That is, we do not know where the global minimum is, but we always know the global minimum is $0$ and all other local traps are strictly positive values. Moreover, the derivative of the energy function $K(\boldsymbol{x})$ is a 3-rd degree polynomial. Define the detection function
\begin{equation}\label{Eqn.6}
L(K) \coloneqq \frac{\Delta K}{K} = \frac{dK}{K\cdot dt}
\end{equation}
and according to L'Hospital's rule, $L(K)$ will approach $0$ when encountering a local trap and will approach $+\infty$ when it is the global minimum. This quantitative determinant could also be used in other research fields when the problem is finding a global minimum of an energy function that is zero only at the global minimum.
In the following Section \ref{Sec.4}, we will give a detailed iterative method and compare its performance with randomization as an escape mechanism.

\section{Local Trap Escape Algorithm}\label{Sec.4}
\subsection{Impulse algorithm}
\noindent If we regard the phase trajectory as a moving point wandering around the 0-1 hypercube, the moving point sometimes will encounter local traps. As mentioned before, the detection function (Equation \eqref{Eqn.6}) can be used for automatically triggering an escape from a local trap by a method we call the \emph{impulse algorithm}.
In Section \ref{results} we test it as a way of escaping local traps and compare it with randomization as an escape mechanism to see which performs better in finding solutions.

The idea of the impulse algorithm is to add another term to Equation \eqref{Eqn.3} so that the trajectory can escape out of a local trap when the detection function (Equation \eqref{Eqn.6}) exceeds a given threshold. First, the Heaviside step function is applied to the detection function to trigger the impulse:
\begin{equation}\label{Eqn.7}
  H[L] =
  \begin{cases}
  1, & L - L_0 \geq 0\\
  0, & \text{otherwise}
  \end{cases}
\end{equation}
where $L_0$ is a threshold and $L$ is the detection function (Equation \eqref{Eqn.6}). Then, an impulse, which is a direction vector $\boldsymbol{I}(K)$ with amplitude $H$, is added to the gradient, so Equation \eqref{Eqn.3} is amended to:
\begin{equation}\label{Eqn.8}
  \frac{\partial{\boldsymbol{x}}}{\partial{t}} = \nabla K(\boldsymbol{x}) + H \cdot \boldsymbol{I}(K)
\end{equation}
where $\nabla K(\boldsymbol{x})$ is just the right-hand side of Equation \eqref{Eqn.3}. If the moving point encounters a local trap, $\nabla K(\boldsymbol{x})$ will be minute and $H$ will fire the impulse $\boldsymbol{I}(K)$ to escape from the local trap.

The method here is to make use of the current gradient of $K(\boldsymbol{x})$ to indicate the escape direction. To prevent a gradient explosion along certain dimensions, a filter process and a scale factor are applied:
\begin{equation}\label{Eqn.9}
  \boldsymbol{I}(K) = k \cdot F^\alpha ~ \text{ins}(\nabla K)
\end{equation}
where $F$ is the filter matrix defined as:
\[F = \begin{bmatrix}
    0.5 & \frac{0.5}{N - 1} & \cdots & \frac{0.5}{N - 1} \\
    \frac{0.5}{N - 1} & 0.5 & \cdots & \frac{0.5}{N - 1} \\
    \vdots & \vdots & \ddots & \vdots \\
    \frac{0.5}{N - 1} & \frac{0.5}{N - 1} & \cdots & 0.5 \\
  \end{bmatrix}\]
where $N$ is the number of variables, $k$ is a scale factor so that $\frac{1}{N}||\boldsymbol{I}(K)||_1 = \frac{1}{2}$
(i.e., $k=\frac{2}{N}||\nabla \boldsymbol(K)||_1$),
and $\alpha$ is an appropriate factor (the minimum integer) so that $||\boldsymbol{I}(K)||_\infty < 1$.
The purpose of $\text{ins}(\cdot)$ is to fix the sign of each element of a vector to make it point to the inside of the 0-1 hypercube,
\begin{equation}\label{Eqn.10}
  \text{ins}(\nabla K(\boldsymbol{x})) \coloneqq -\text{sgn}[\nabla K] \circ \text{sgn}\left[\boldsymbol{x} - [0.5, 0.5, \cdots, 0.5]^T\right] \circ \nabla K
\end{equation}
where $\circ$ represents the Hadamard (component-wise) product operation. For instance, if $\boldsymbol{x} = [0.1, 0.9]^T$ and $\nabla K(\boldsymbol{x}) = [0.5, 0.6]^T$, then $\text{sgn}[\nabla K] = [1, 1]^T$ and $\text{sgn}\left[\boldsymbol{x} - [0.5, 0.5]^T\right] = [-1, 1]^T$, so $-\text{sgn}[\nabla K] \circ \text{sgn}\left[\boldsymbol{x} - [0.5, 0.5]^T\right] = [1, -1]^T$, which is able to flip the direction of the second variable of $\nabla K$. Although the formula here looks complicated, it is easier to implement it in the computer program.

In brief, $\boldsymbol{I}(K)$ preserves the property of the gradient and scales the amplitude so that the average amplitude across all dimensions is $1/2$. Further, the amplitude along each dimension is less than 1 and the overall direction is toward the interior of the hypercube. This design is intended to make full use of the current local trap information to escape and to find the global minimum with greater probability. For more details, see Section \ref{Sec.4.2}.

To demonstrate this algorithm's effectiveness, we conducted a series of comparisons on different problem sizes for the impulse algorithm and compared it to a randomization method that randomizes all variables when encountering a local trap. The Euler method with step length $1$ was used for the gradient descent. $(M \in \{1, 2, 3, 5, 8, 10, 15\}) \times (N \in \{3, 5, 8, 10, 12, 15\}) \times (R \in \{1, 2, 3, 5, 10, 15\})$ resulted in $7 \times 6 \times 6 = 252$ conditions in total, where $R$ is the range of coefficients (e.g. $R = 5$ means all of the coefficients $0 \leq c_i \leq 5$). For each condition, 200 trials were conducted, and in each trial the maximum number of iterations was 1000. We used a threshold $L_0 = 10^{-4}$ but the value is not critical.

\subsection{Results}\label{results}
\label{Sec.4.2}
\begin{figure}[htbp]
  \centering
  \subfigure[]{
    \includegraphics[width=0.45\textwidth]{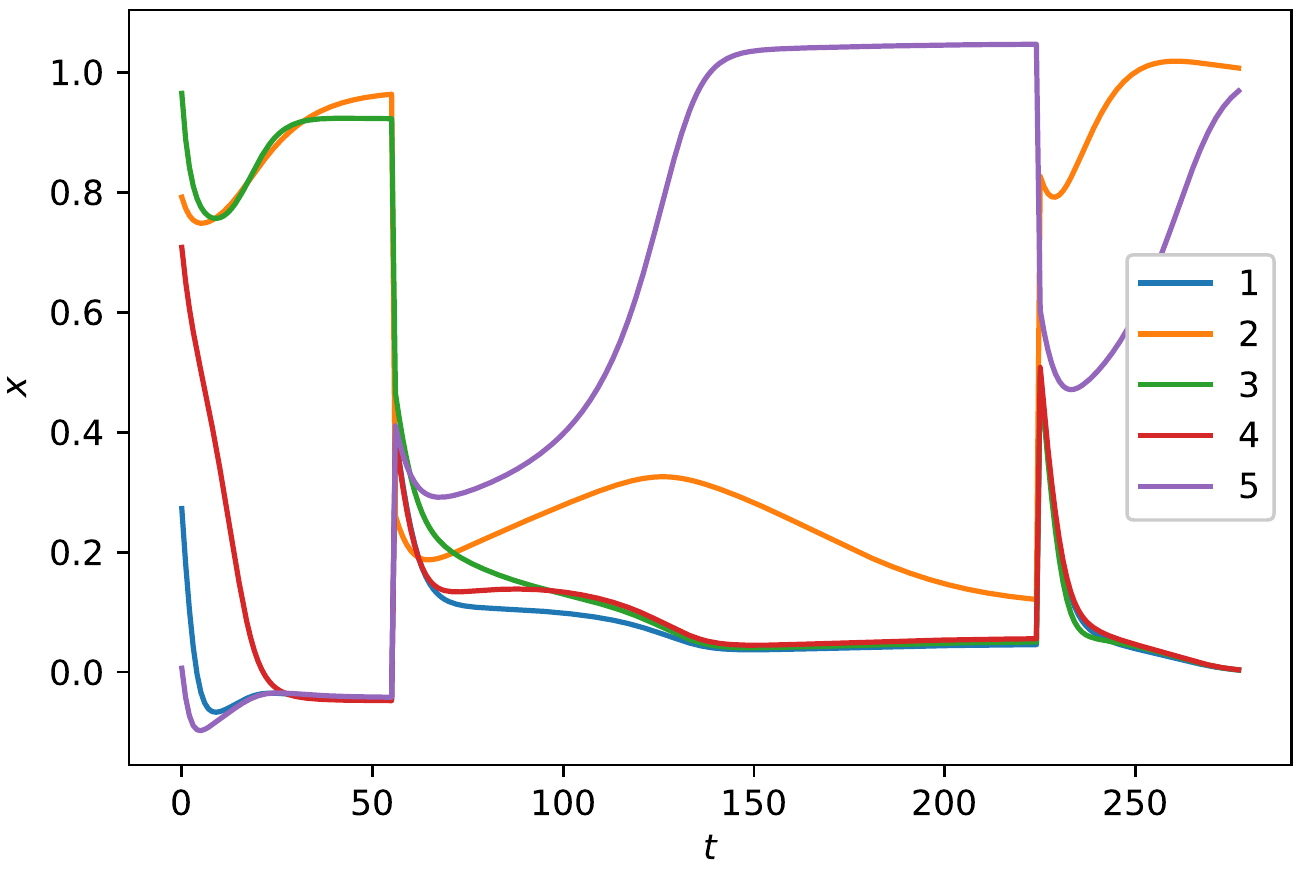}
    \label{Fig.2.sub.a}}
  \subfigure[]{
    \includegraphics[width=0.45\textwidth]{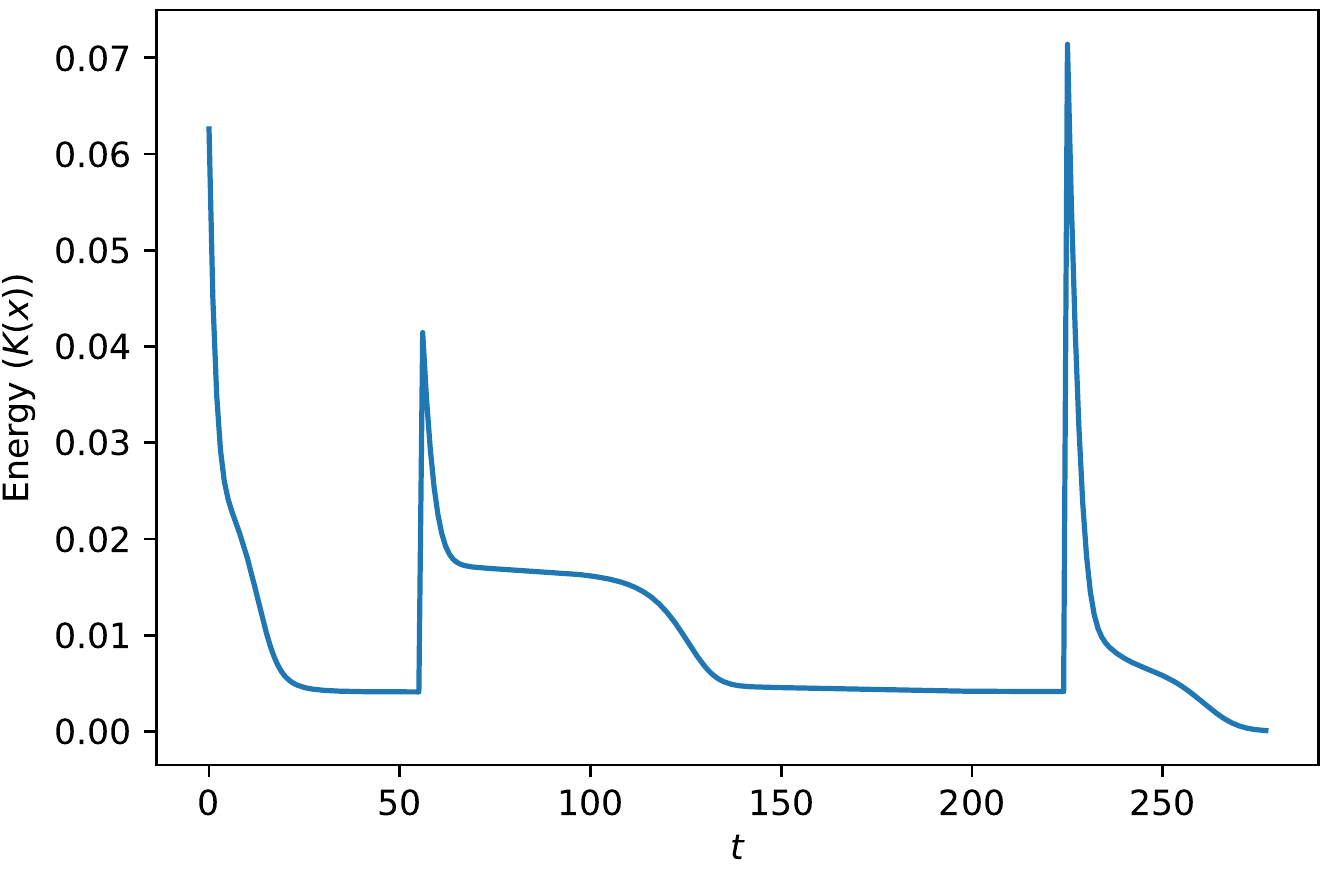}
    \label{Fig.2.sub.b}}

  \subfigure[]{
    \includegraphics[width=0.45\textwidth]{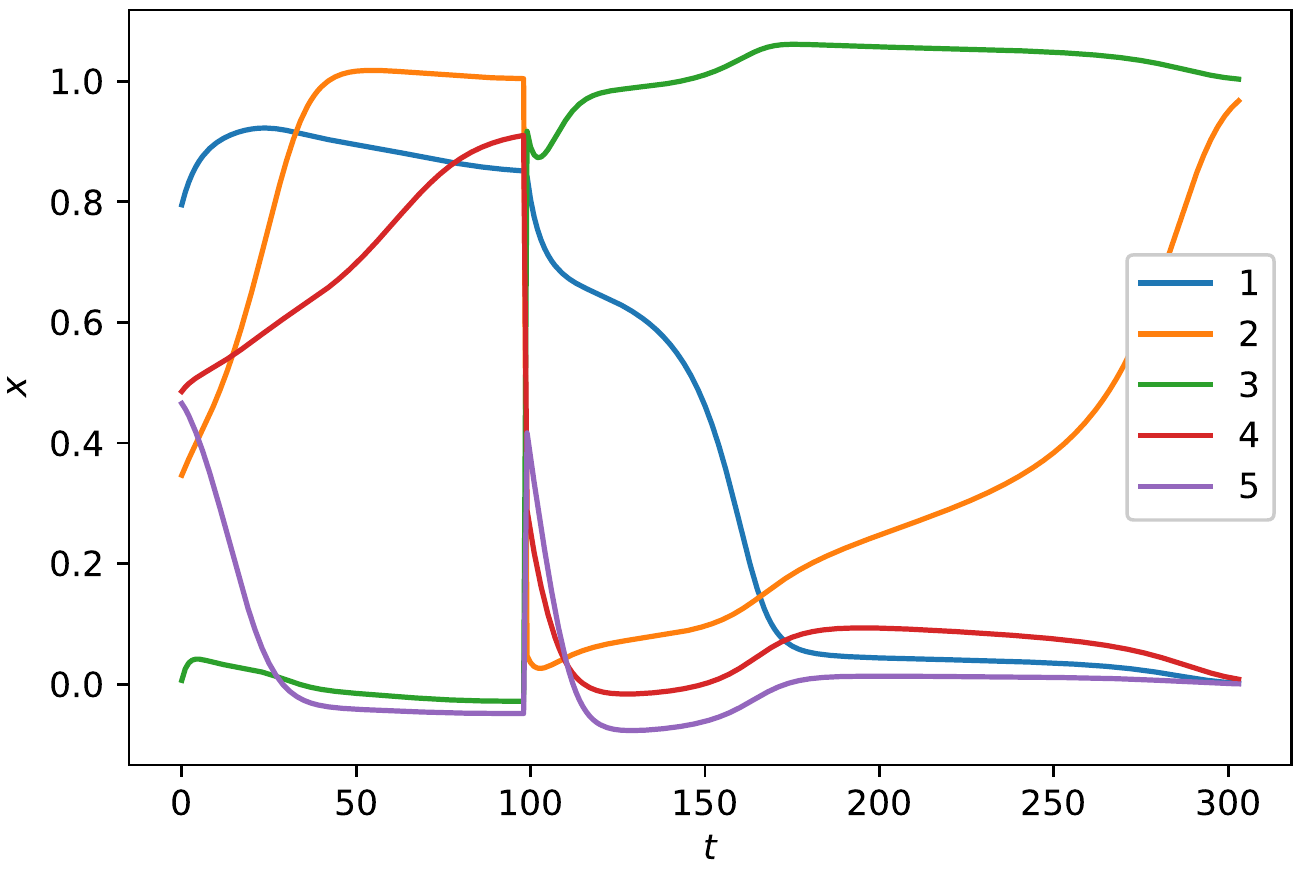}
    \label{Fig.2.sub.c}}
  \subfigure[]{
    \includegraphics[width=0.45\textwidth]{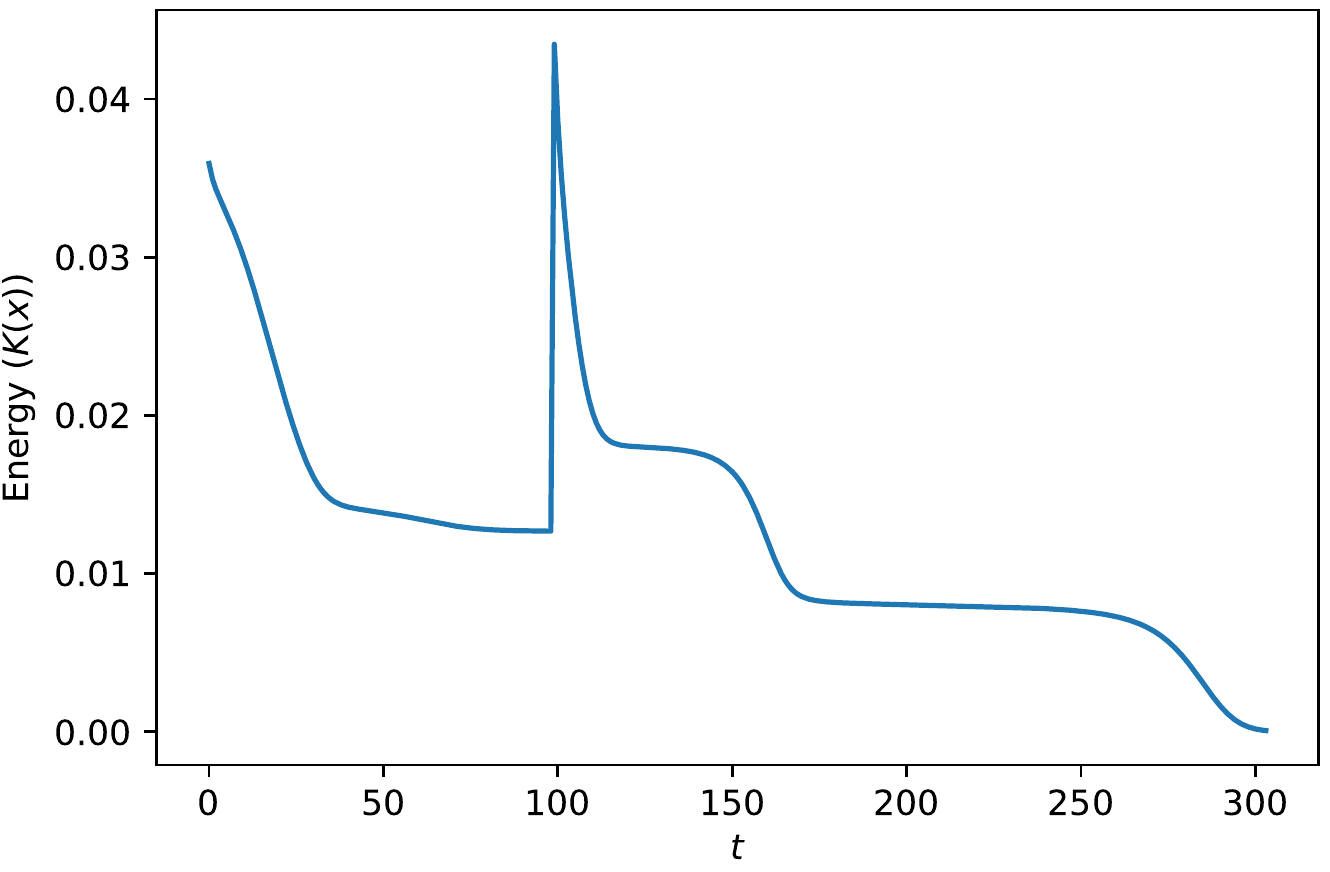}
    \label{Fig.2.sub.d}}
  \caption{Running examples of impulse algorithm (a, b) and randomization method (c, d). $M = 3$, $N = 5$, $R = 10$. (a, c) represent the moving point, (b, d) represent the corresponding energy.}
  \label{Fig.2}
\end{figure}

\noindent Running examples are shown in Figure \ref{Fig.2}. When the trajectory has fallen into a local trap, the impulse algorithm makes use of the current gradient information to fire an escape impulse directed to the interior of the hypercube (in Figure \ref{Fig.2.sub.a}, two pulses appear at about $t = 60$ and $t = 230$), whereas the randomization method generates a random vector with components uniformly distributed between 0 and 1.

\begin{figure}[htbp]
  \centering
  \subfigure[]{
    \includegraphics[width=0.16\textwidth]{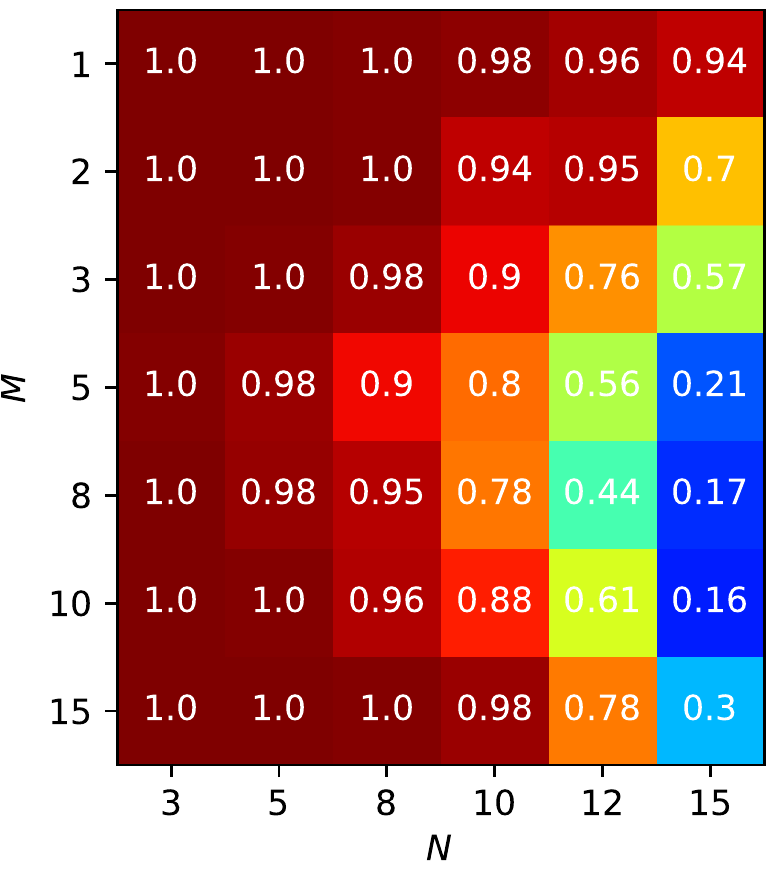}
    \includegraphics[width=0.16\textwidth]{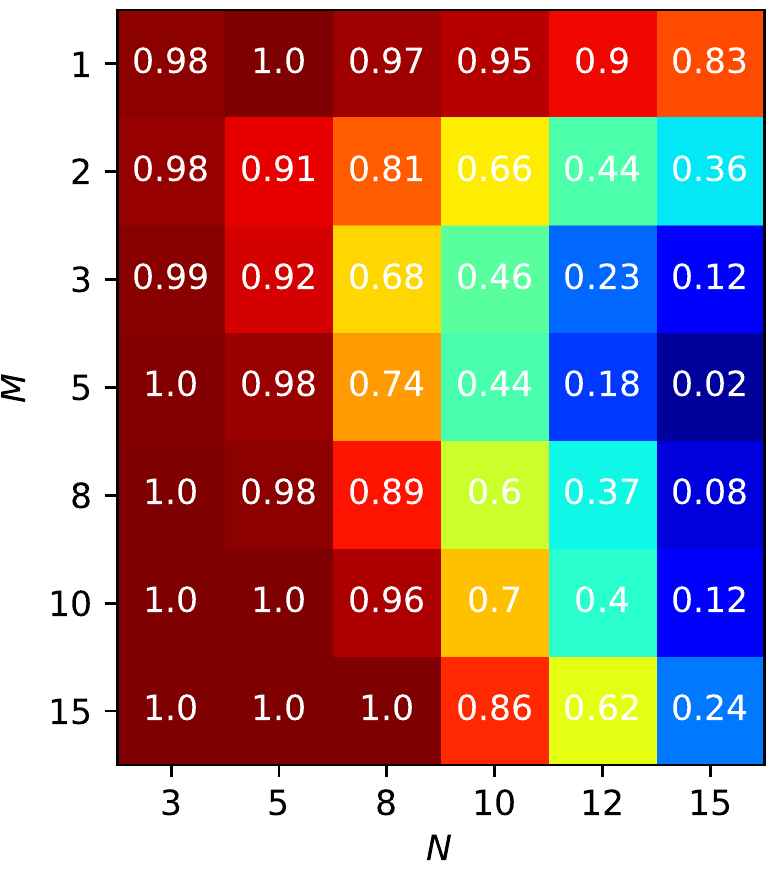}
    \includegraphics[width=0.16\textwidth]{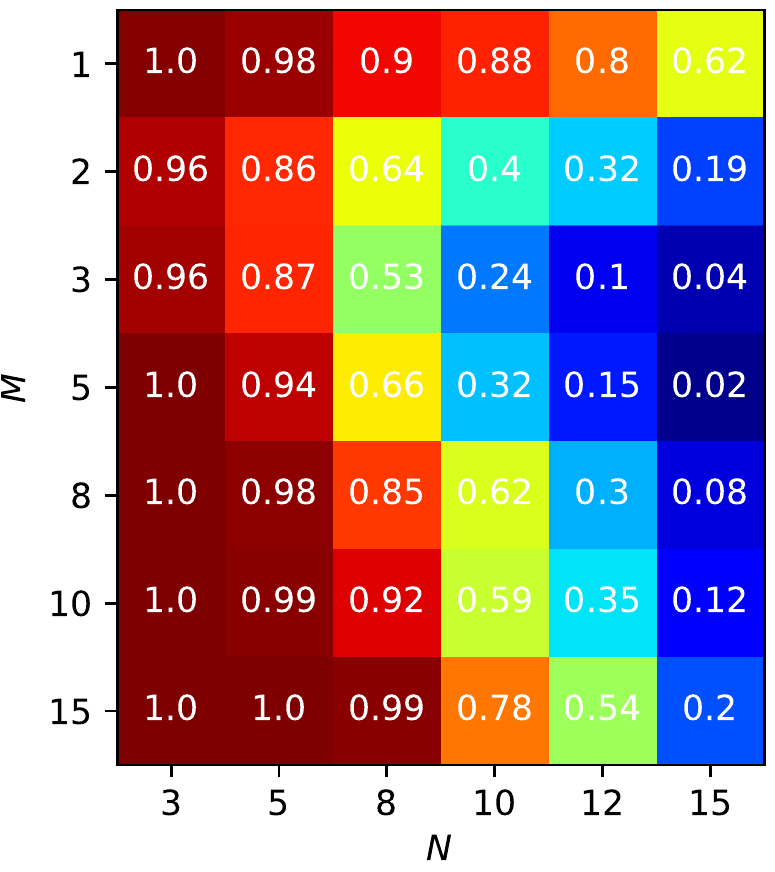}
    \includegraphics[width=0.16\textwidth]{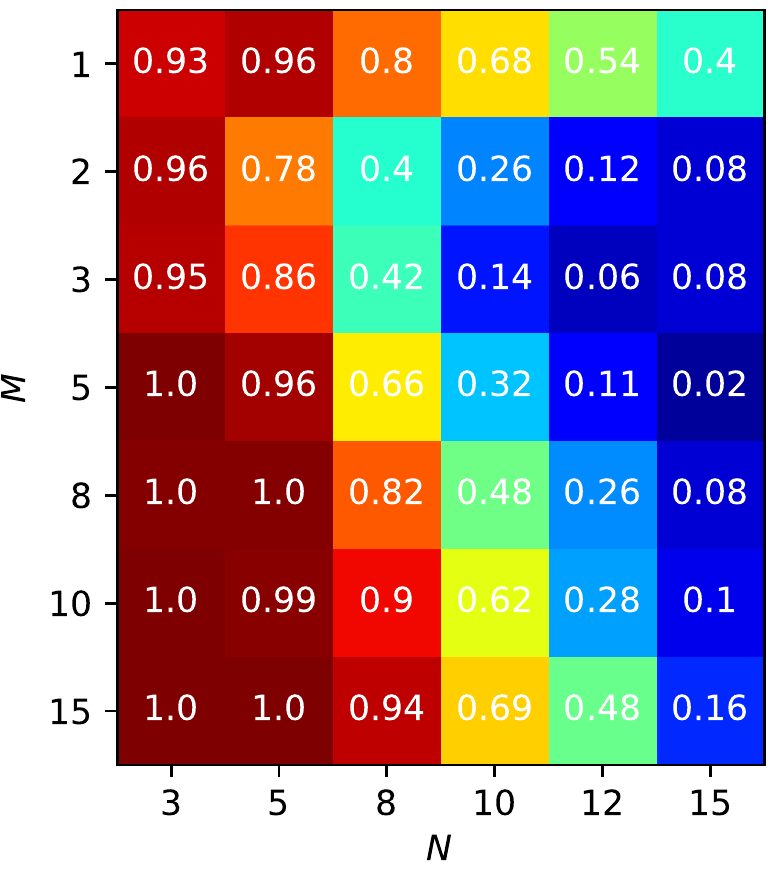}
    \includegraphics[width=0.16\textwidth]{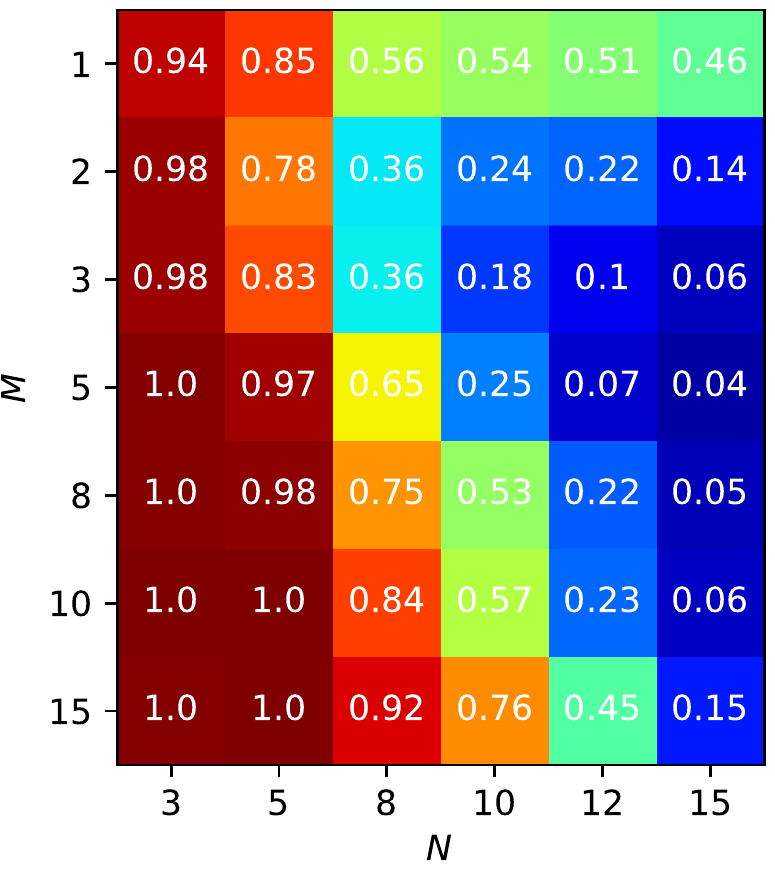}
    \includegraphics[width=0.16\textwidth]{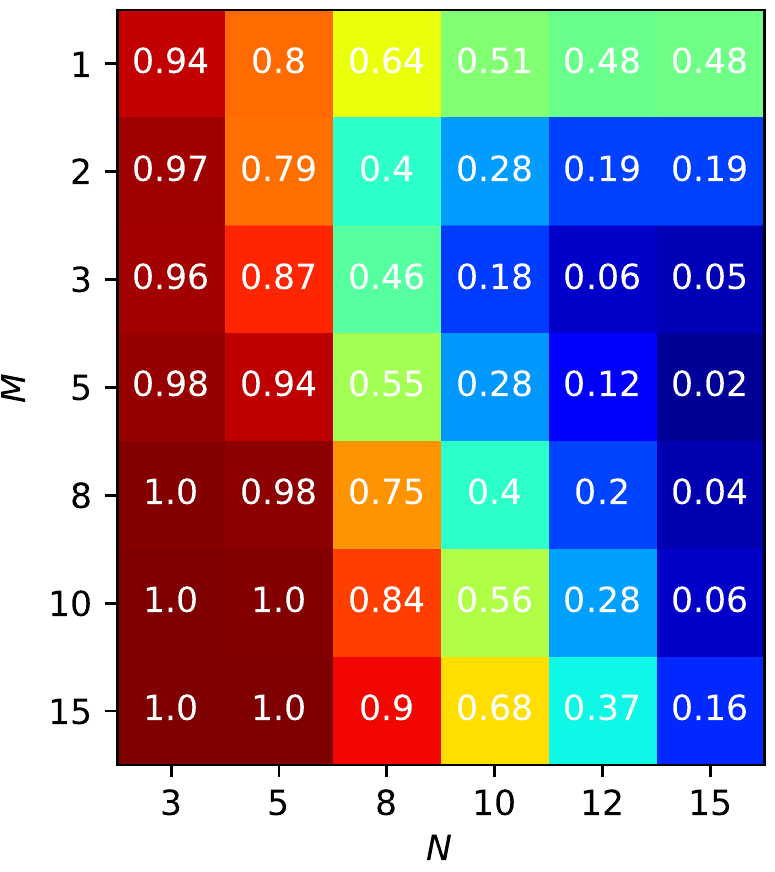}
    \label{Fig.3.sub.a}}

  \subfigure[]{
    \includegraphics[width=0.16\textwidth]{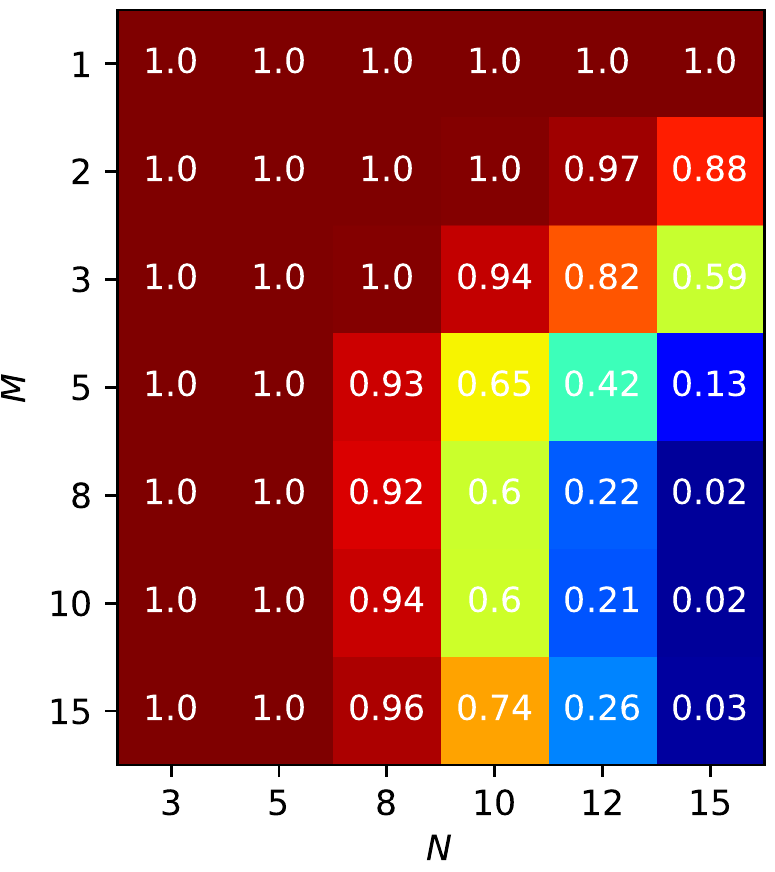}
    \includegraphics[width=0.16\textwidth]{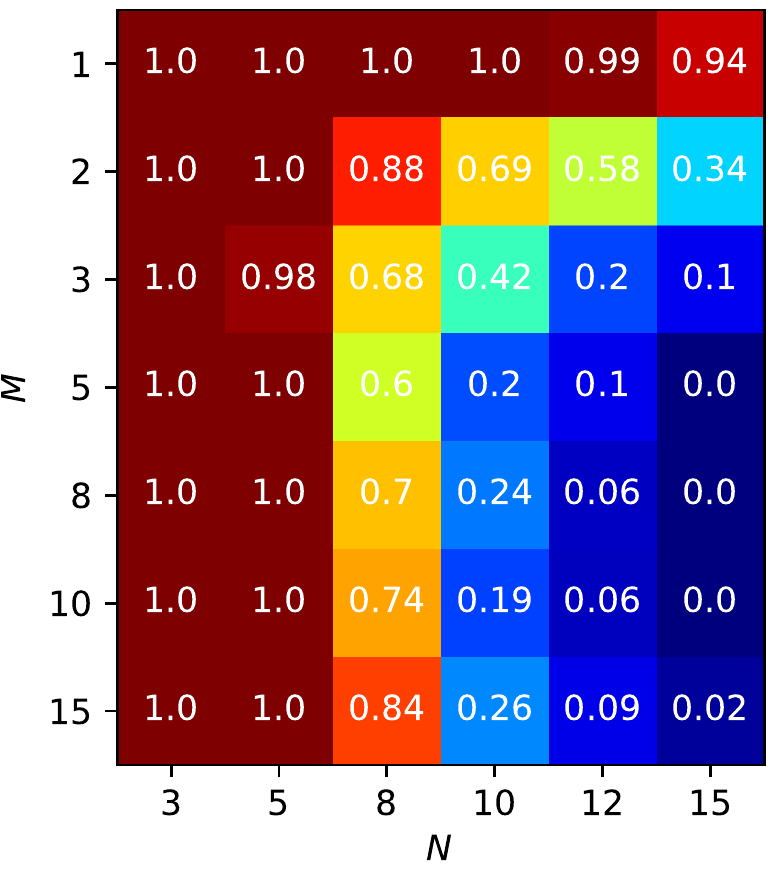}
    \includegraphics[width=0.16\textwidth]{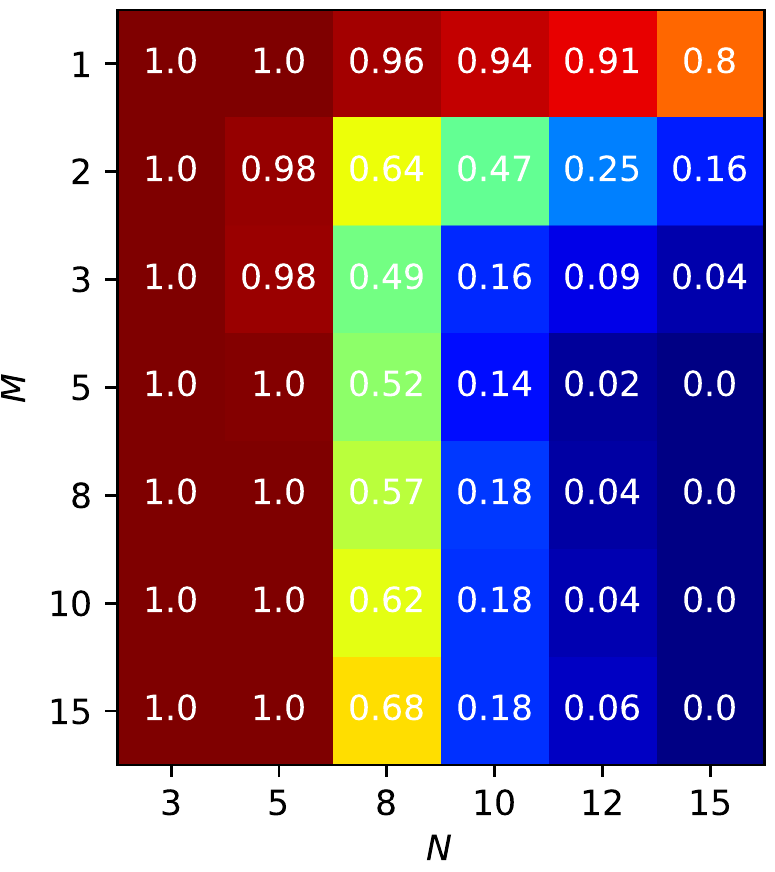}
    \includegraphics[width=0.16\textwidth]{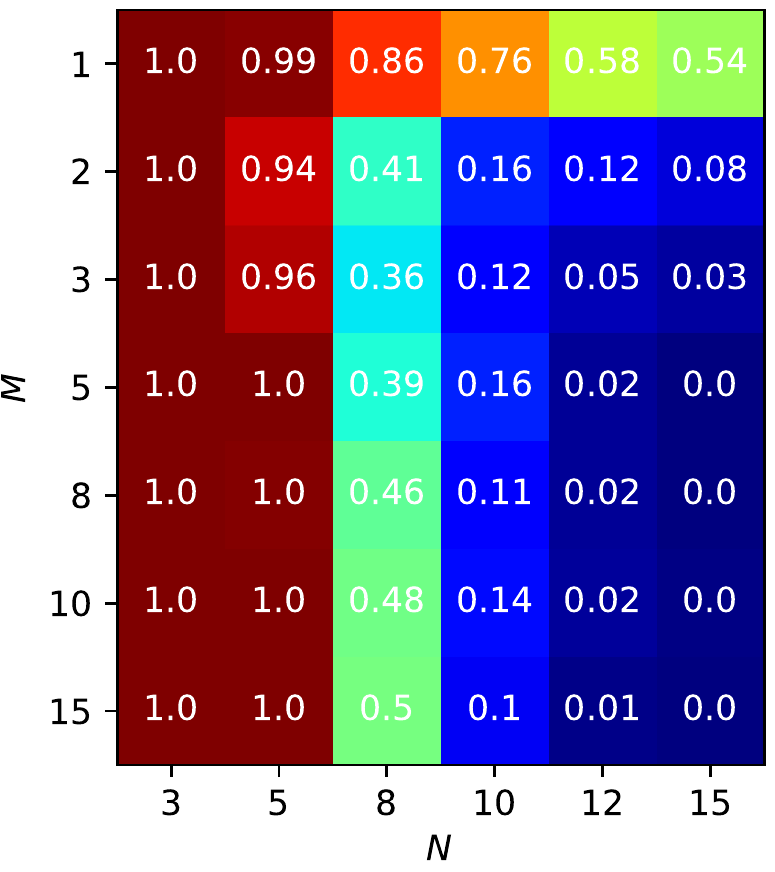}
    \includegraphics[width=0.16\textwidth]{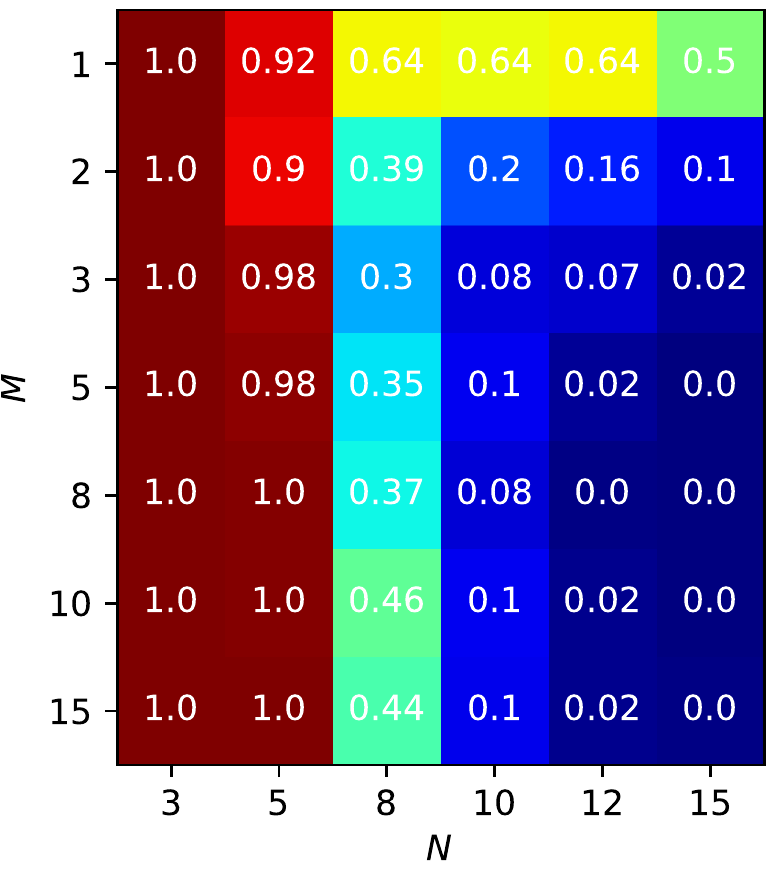}
    \includegraphics[width=0.16\textwidth]{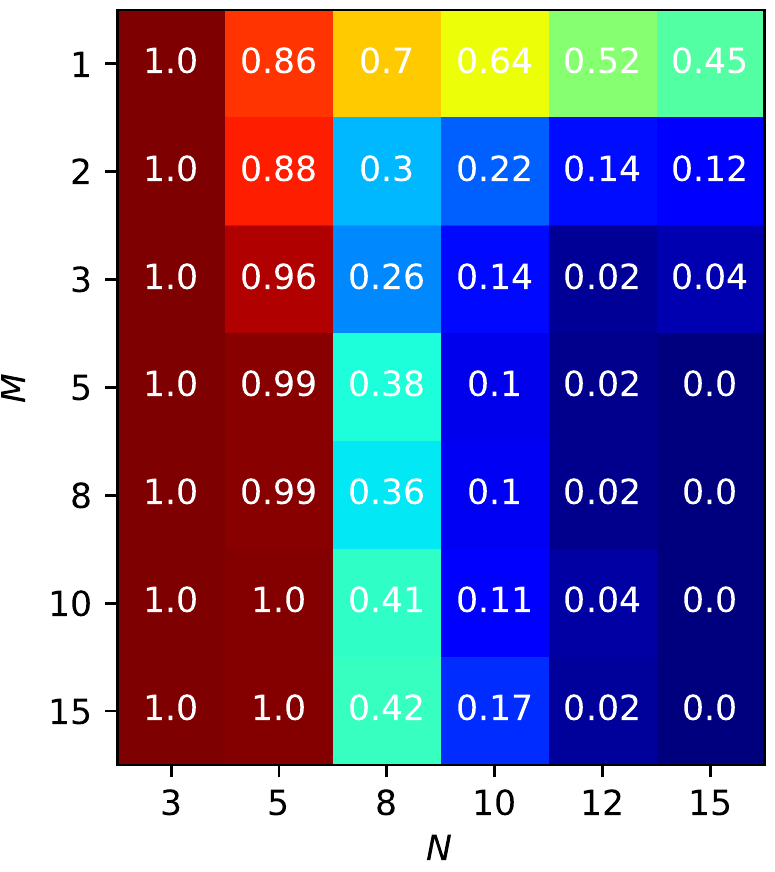}
    \label{Fig.3.sub.b}}

  \caption{Success rates of escape algorithm (a) and randomization method (b). From left to right, $R = 1, 2, 3, 5, 10, 15$.}
  \label{Fig.3}
\end{figure}

Figure \ref{Fig.3} compares the results. In general, the success rate decreases with increasing $N$ and $R$, and increases with increasing $M$, except for $M = 1$ since there may be more than one solution to a specific problem. This phenomenon becomes rare when the density $M/N$ is relatively large and the coefficient range $R$ is large. In this comparison, the impulse algorithm is obviously better than the randomization method in the $N > 5$ conditions. For instances, under the conditions $M > 5, N = 15$, it is almost impossible for the randomization method to find the global minimum in 1000 steps, but the impulse method succeeded in several trials out of 200 total trials.

To understand why the impulse algorithm performs better, we may go back to the energy function $K(\boldsymbol{x})$ (Equation \eqref{Eqn.2}). $K$ is a concave-up fourth-order polynomial. Along each dimension, there are two minima, and one of them is the global minimum located exactly at 0 or 1. Given that the trajectory is initiated within the 0-1 hypercube, under normal circumstances it would only move around the hypercube (as long as the step length is carefully controlled).
Thus, because of the effect of the first term on the second term of $K$, the local minimum along some dimensions might be repelled far away from the hypercube, so that the moving point will not drop into that local trap. Furthermore, the energy surface along those dimensions are usually steep, so its directional derivative in those dimensions approaches 0 quickly, and this leads to the large escape occurring mainly along other dimensions, because of the property of the iterative formula \eqref{Eqn.8} of the impulse algorithm.

\section{Time-to-solution Distribution of the Impulse Algorithm}\label{Sec.5}

\noindent Since the impulse algorithm does not include any random process during the search process, it is interesting to see the time-to-solution distribution for some particular sized problems. Hence, we extend the maximum number of iterations to see the time-to-solution distribution for $(M \in \{3, 5\}) \times (N \in \{5, 8, 10\}) \times (R = 10)$. Each condition has 500 trials.

\begin{figure}[htbp]
  \centering
  \subfigure[]{
    \includegraphics[width=0.3\textwidth]{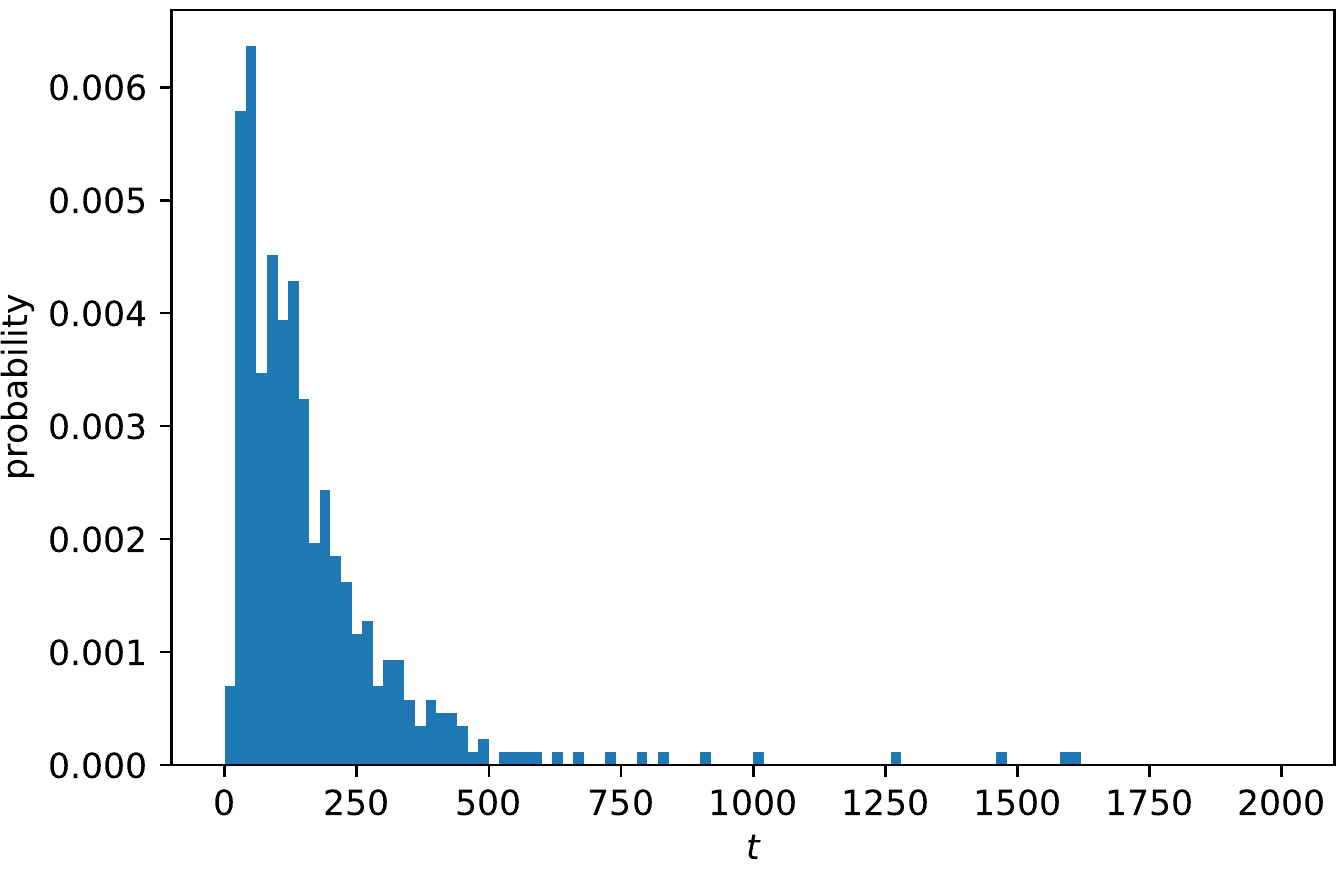}
    \includegraphics[width=0.3\textwidth]{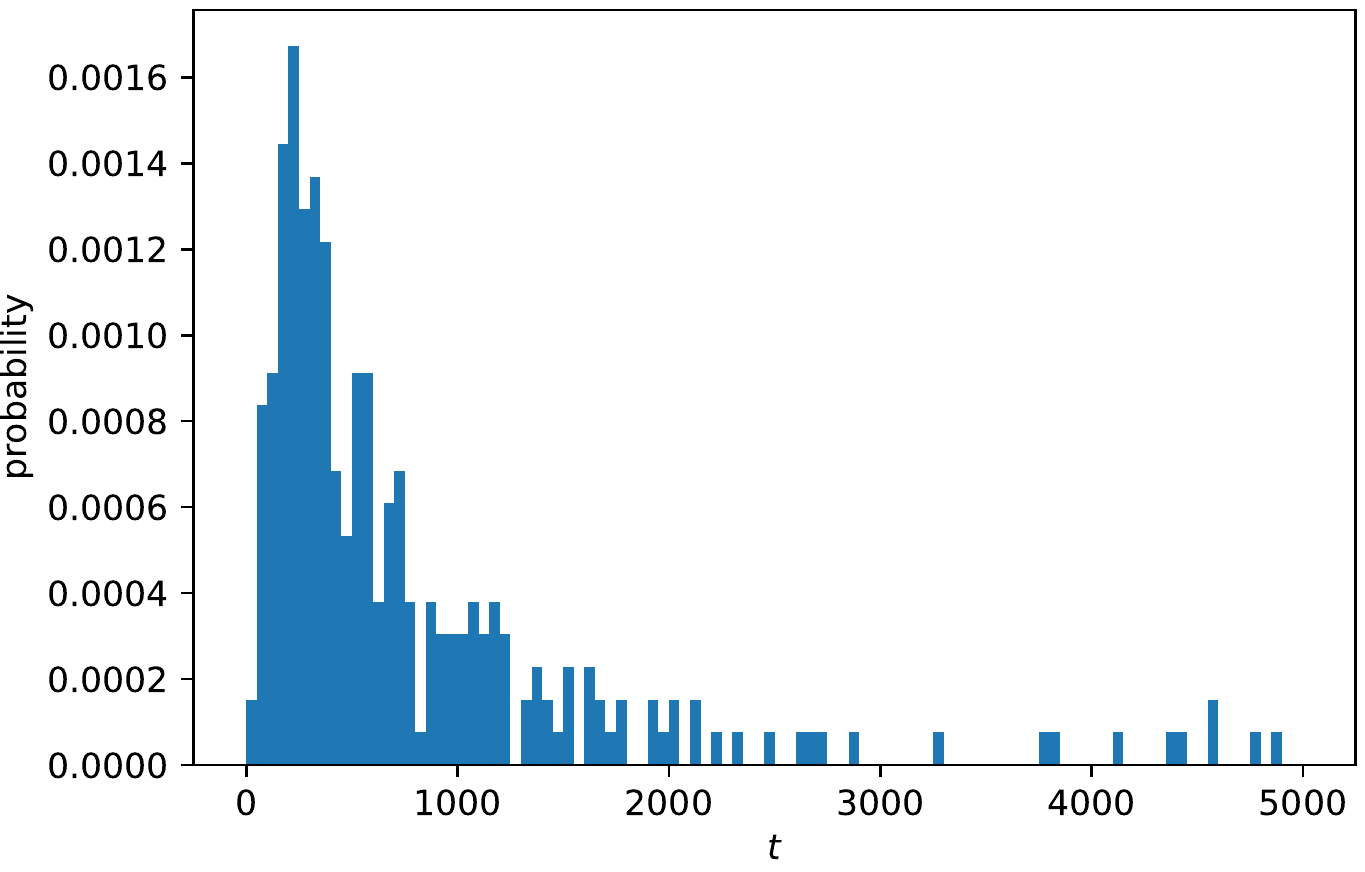}
    \includegraphics[width=0.3\textwidth]{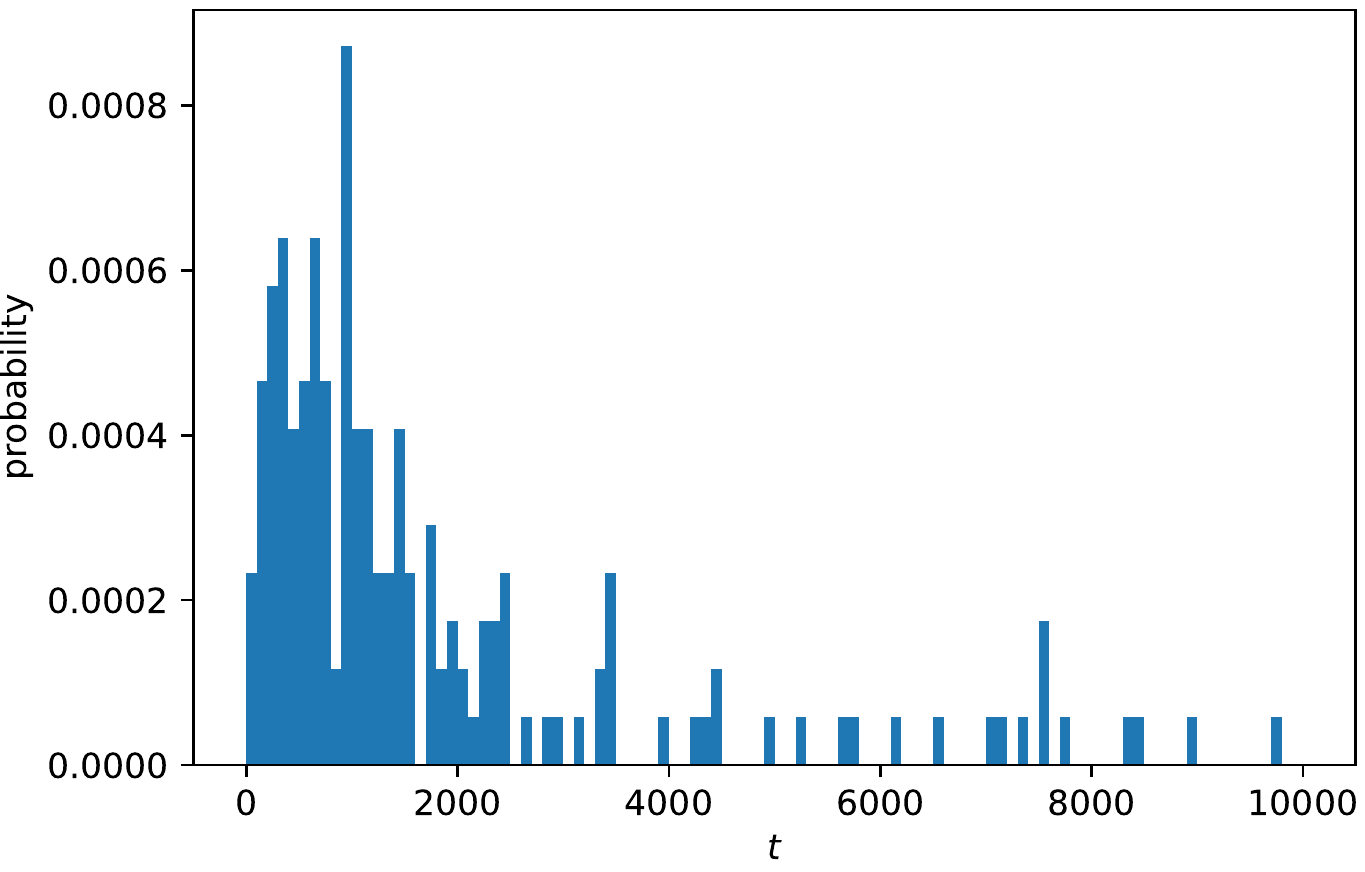}
    \label{Fig.4.sub.a}}

  \subfigure[]{
    \includegraphics[width=0.3\textwidth]{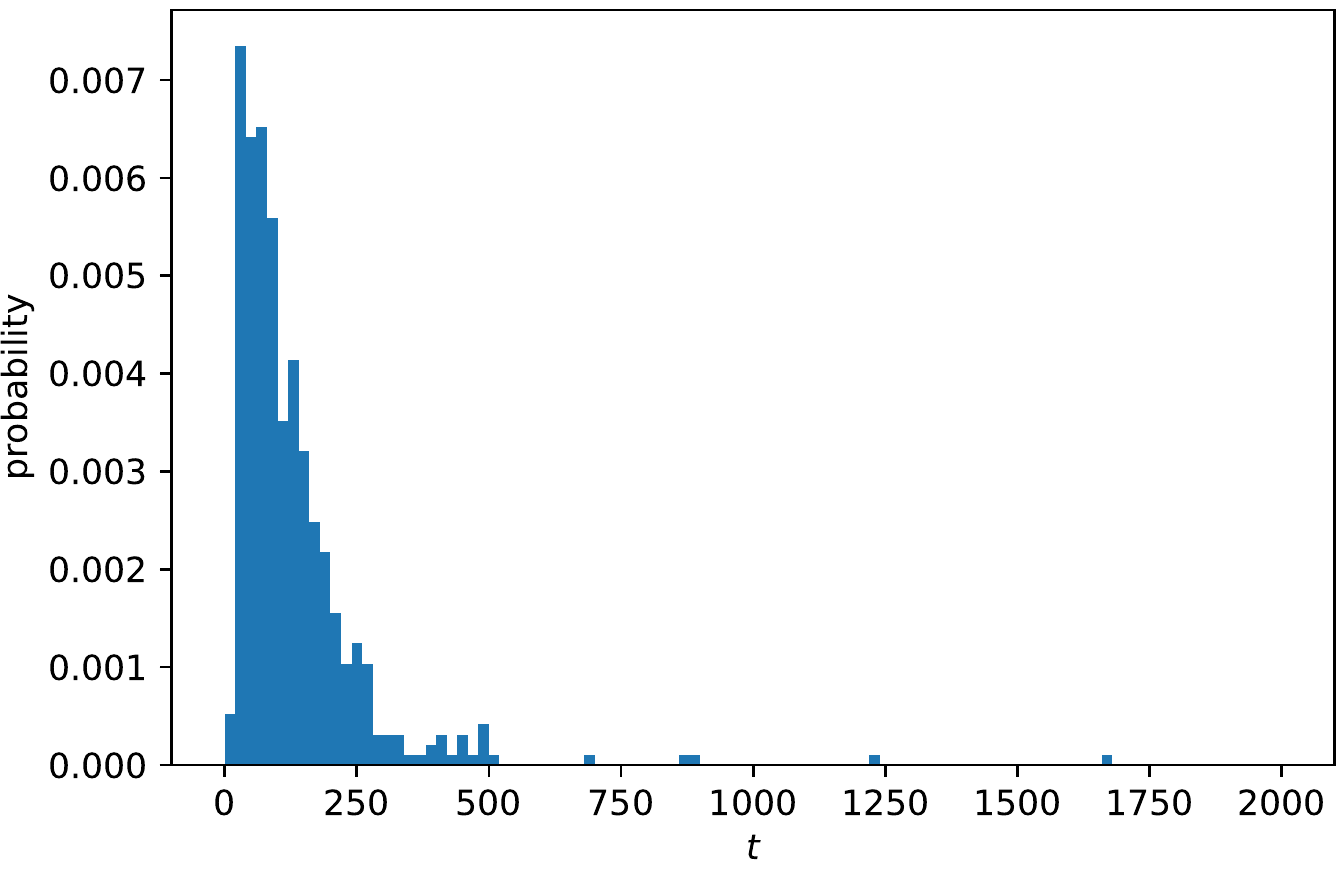}
    \includegraphics[width=0.3\textwidth]{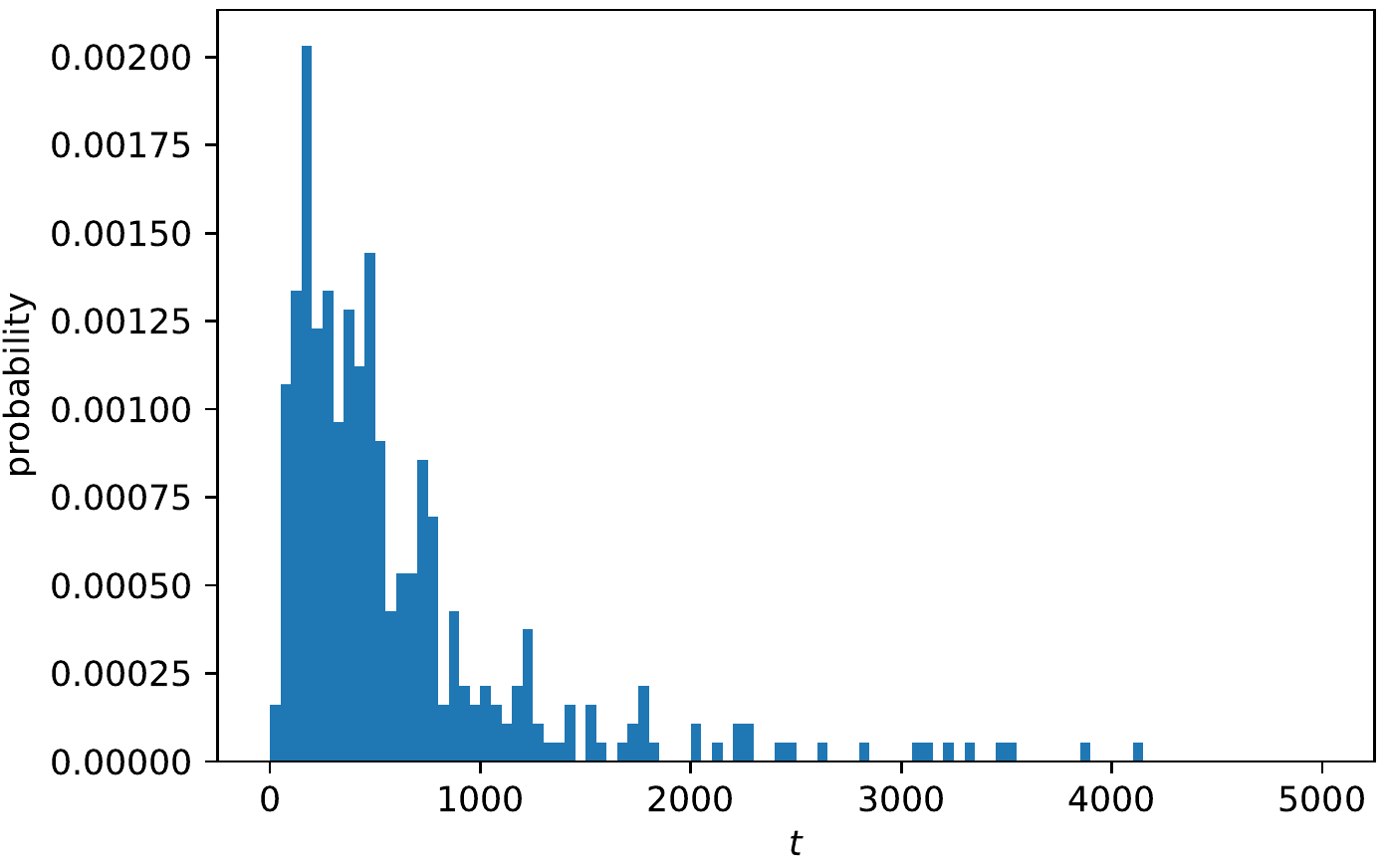}
    \includegraphics[width=0.3\textwidth]{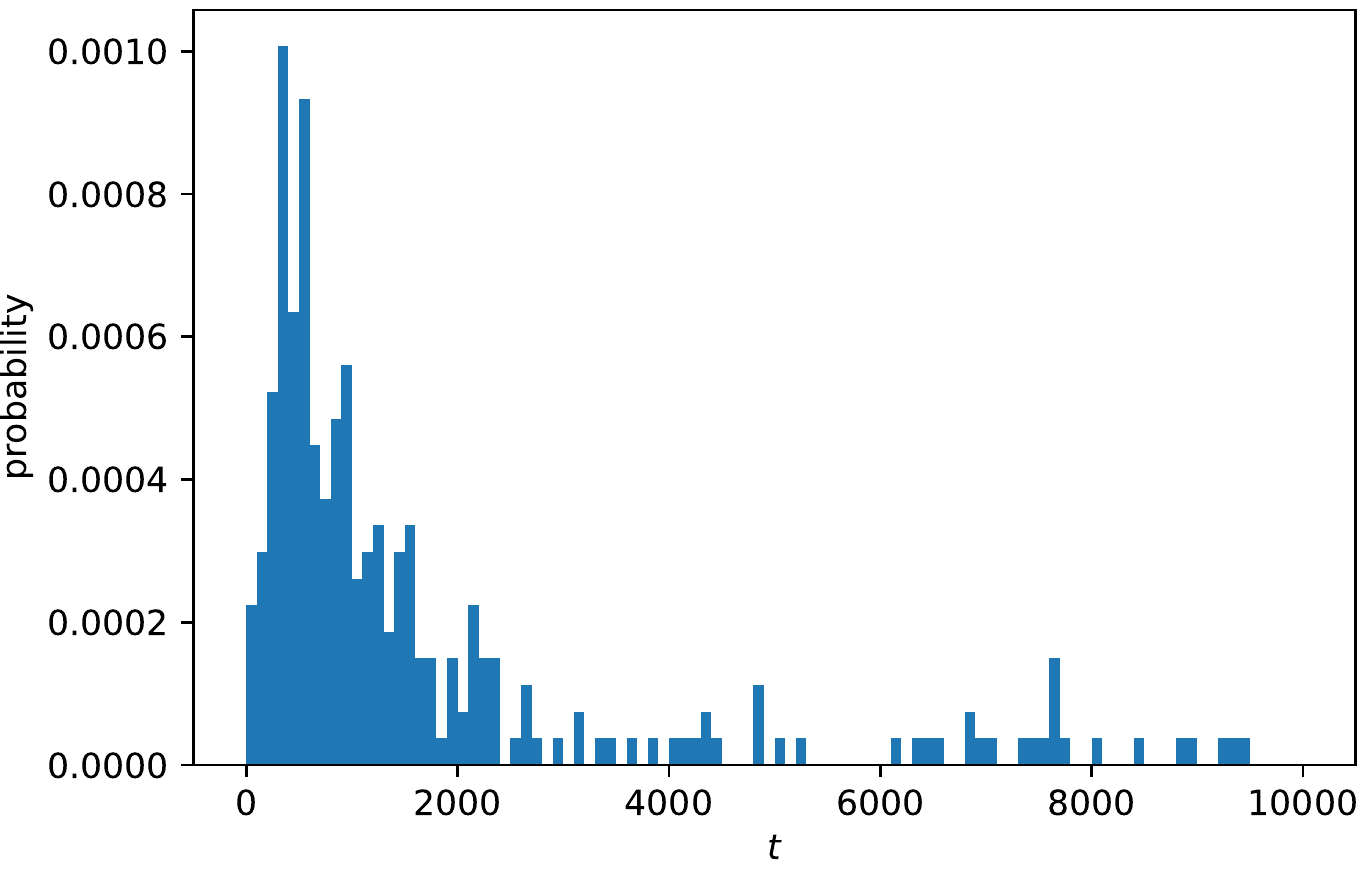}
    \label{Fig.4.sub.b}}

  \caption{Time-to-solution distribution for $M = 3$ (a) and $M = 5$ (b). Left: $N = 5$; middle: $N = 8$; right: $N = 10$. 100 bins for all conditions. Trials not finished in 2000 iterations for $N = 5$, 5000 iterations for $N = 8$, and 10000 iterations for $N = 10$ are not shown in this figure.}
  \label{Fig.4}
\end{figure}

\begin{figure}[htbp]
  \centering
  \includegraphics[width=0.5\textwidth]{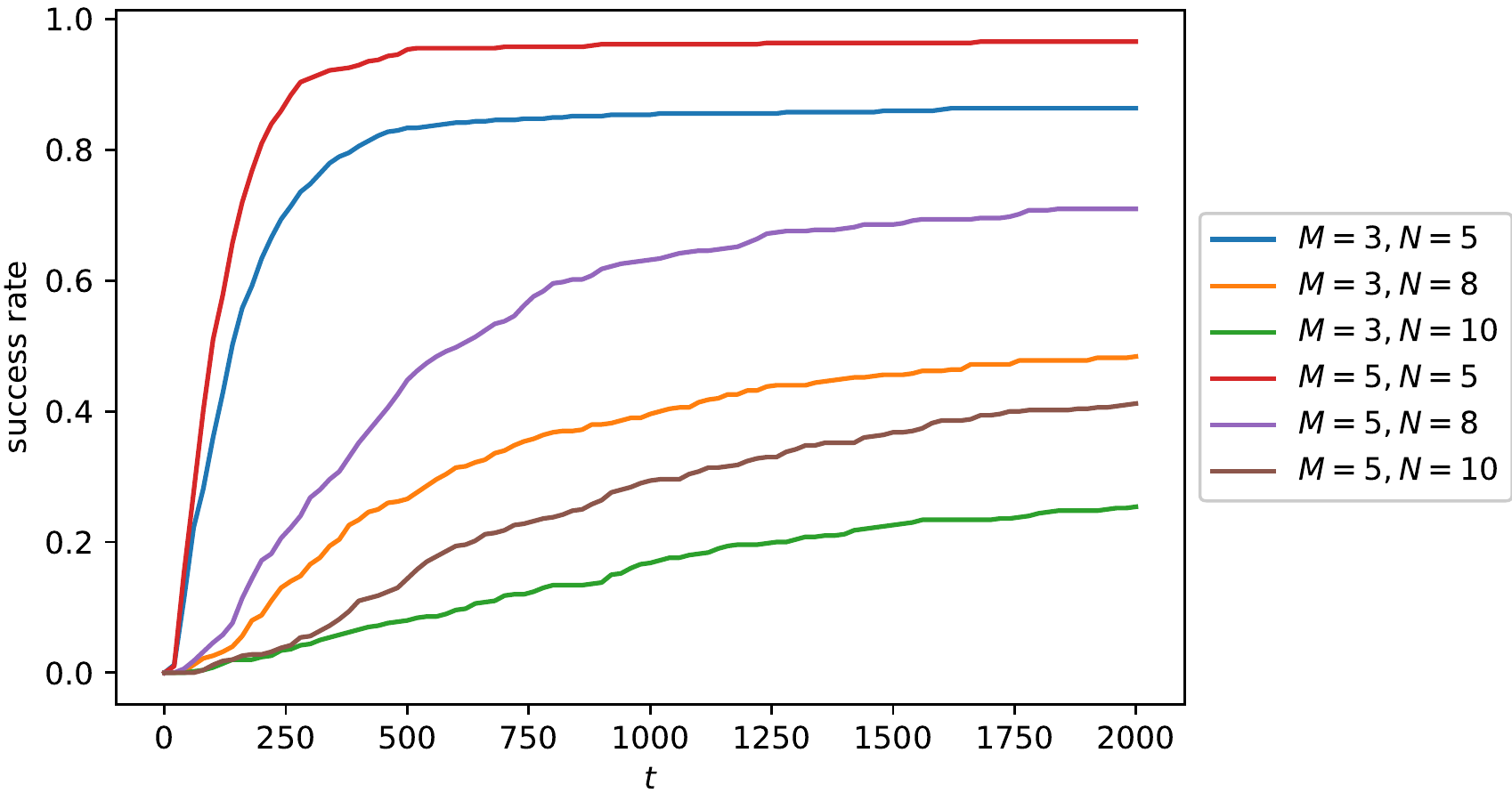}
  \caption{Comparison of cumulative success rates as a function of time $t$ within 2000 iterations. Coefficient range $R = 10$.}
  \label{Fig.5}
\end{figure}

The time-to-solution distribution results (Figure \ref{Fig.4}) are consistent with the results in the previous section. With increasing  $N$, the time-to-solution becomes longer. The greater the density ($M/N$) of the problem, the less time it takes to find the solution. An explicit comparison of these six conditions is shown in Figure \ref{Fig.5}. The simpler the problem, the higher the cumulative success rate and the sooner it tends to stabilize.

It is worth mentioning that the time-to-solution distribution is not a uniform distribution at all. These results indicate that most of the instances can be solved quickly by the impulse algorithm and only a few of them are difficult. For example, the median time of the condition $M = 3, N = 5$ is $138.5$ and the average time of that condition is larger than $417.8$ (since there are 17 instances not yet solved by $t = 2000$). These statistics mean that more than half of the instances can be solved below the average time. When it comes to the exhaustive search method, however, its time-to-solution distribution is a uniform distribution. Therefore, in average cases, the impulse algorithm is more likely to find a solution in a short time compared with the exhaustive search method.

\section{Basin of Attraction Estimation}\label{Sec.6}
\subsection{Method}
\noindent As discussed in Section \ref{Sec.4.2}, we speculate that the number of minima around the 0-1 hypercube is less than $2^N$. If we treat the randomized initial position of the moving point as casting the point on the energy surface, we also hypothesize that the basin of attraction of the global minimum could be larger than $2^{-N}$, the discrete probability in an exhaustive search.
(The basin of attraction of a minimum is the high dimensional region in which all trajectories to descend to that minimum.) 

In case there could be more than one solution for some particular sized problems, we conducted 100 trials for each condition of $(M \in \{3, 5, 10\}) \times (N \in \{5, 8, 10, 12, 15\}) \times (R \in \{3, 5, 10\})$ to see if the attracting area is larger than $2^{-N}$. Of course, it is almost impossible to have multiple solutions in these conditions. In each trial, 100 points are randomly initialized within the 0-1 hypercube to calculate how many points are in the global minimum's basin of attraction. The basin of attraction for a particular condition is estimated by averaging across the 100 trials in that condition to get an approximation of the basin of attraction of the global minimum.

\subsection{Results}

\noindent The comparison is shown in Table \ref{Tab.1}. It is obvious that all of the basins of attraction for these conditions are at least ten times larger than $2^{-N}$ (the discrete probability). This means the basin is quite large in the 0-1 hypercube. As discussed before, this may happen because some local traps are far from the 0-1 hypercube, or because the energy surface itself has a larger potential well for the global minimum.

\begin{table}[htbp]
  \label{Tab.1}
  \centering
  \caption{Comparison table between the global minimum's basin of attraction and the discrete probability $2^{-N}$. Each column represents the ratio of the basin area to $2^{-N}$. For each $M$ and $N$, the three values represent the conditions $R = 3, 5, 10$ respectively.}
  \begin{tabular}{cccccc}
  \toprule
  $M \backslash N$ & 5 & 8 & 10 & 12 & 15 \\
  \midrule
  3 & 14.5, 12.7, 10.1 & 26.4, 16.3, 14.1 & 38.2 , 25.9, 20.5 & 79.1, 27.9, 34.8 & 268.7, 258.9, 203.2 \\
  5 & 18.1, 15.0, 13.7 & 25.3, 20.2, 11.9 & 24.6, 13.5, 15.9 & 24.2, 14.8, 32.8 & 19.7, 59.0, 6.6 \\
  10 & 24.5, 20.7, 16.6 & 41.3, 26.4, 17.5 & 35.3, 26.4, 15.3 & 18.8, 28.7, 16.4 & 16.4, 16.4, 3.3 \\
  \bottomrule
 \end{tabular}
\end{table}

When the coefficient range becomes larger (from $[0, 3]$ to $[0, 10]$), the basin size shows a decreasing trend, except for the conditions $M = 5, N = 10$; $M = 5, N = 15$; and $N = 12$. This suggests that the narrower the coefficient range is, the simpler the problem is, and the simpler the energy function surface is. Therefore, compared with the exhaustive search method, the impulse algorithm has a greater advantage for the narrower coefficient range. However, with respect to $M$ and $N$, the ratio of the basin area does not have an obvious uniform trend. Besides, it is interesting that for the condition of $M = 3, N = 15$, the ratios are all greater than 200, which are significantly greater than other conditions.

\section{Locating the Basin of Attraction}\label{Sec.7}
\subsection{Method}
\noindent Since the results in the previous section have proved the global minimum possesses a relatively larger basin of attraction, we were curious about where these basins are and whether there is any relationship between its corresponding constraint matrix $C$ and object vector $\boldsymbol{d}$. Here, we conduct a more detailed point casting experiment only for the condition $M = 3, N = 10, R = 10$. It contains 200 trials (200 different problems) and 5000 points are randomly cast into the 0-1 hypercube in each trial. Initial positions and their corresponding constraint matrix $C$ and object vector $\boldsymbol{d}$ in each trial are recorded.

\subsection{Results}
\begin{figure}[htbp]
  \centering
  \includegraphics[width=0.45\textwidth]{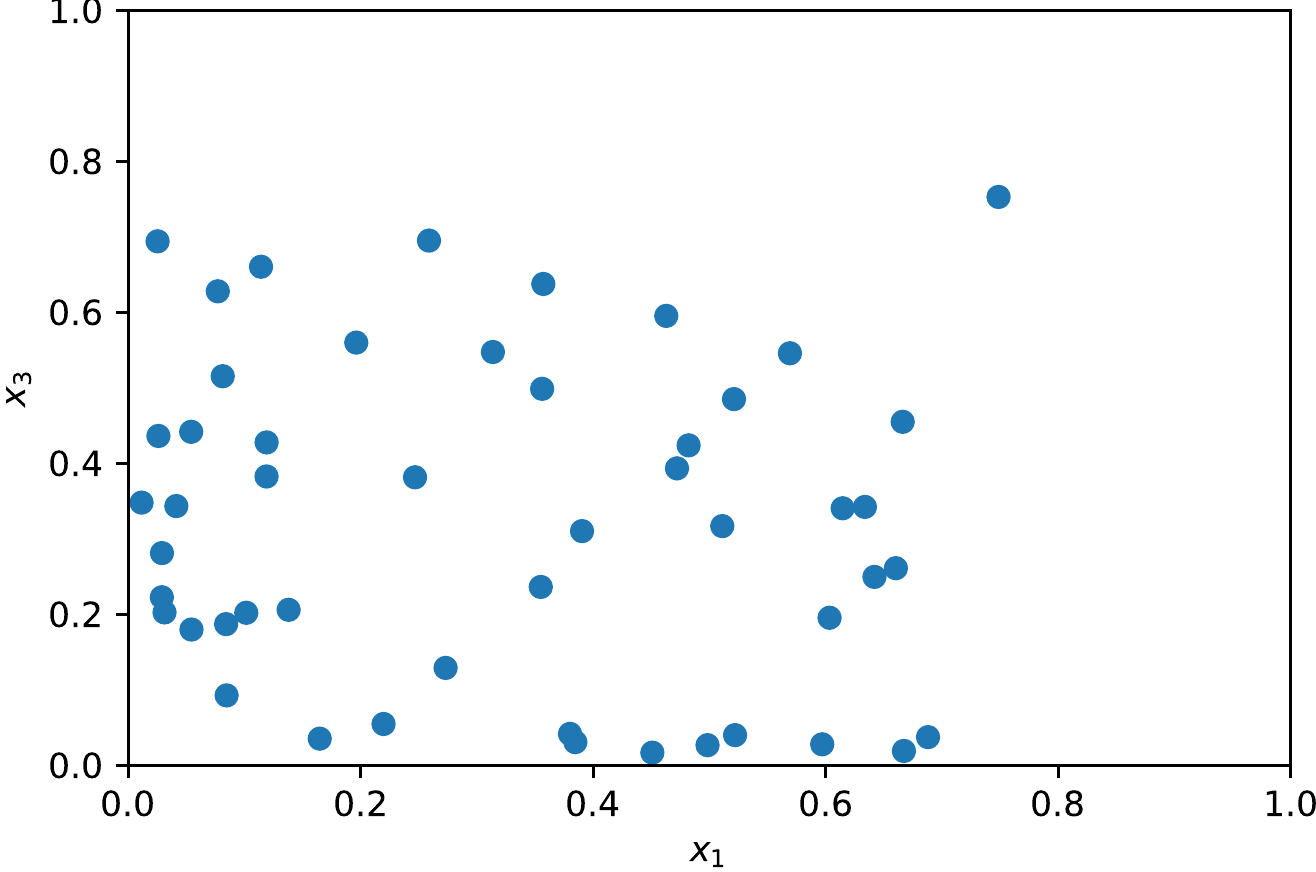}
  \includegraphics[width=0.45\textwidth]{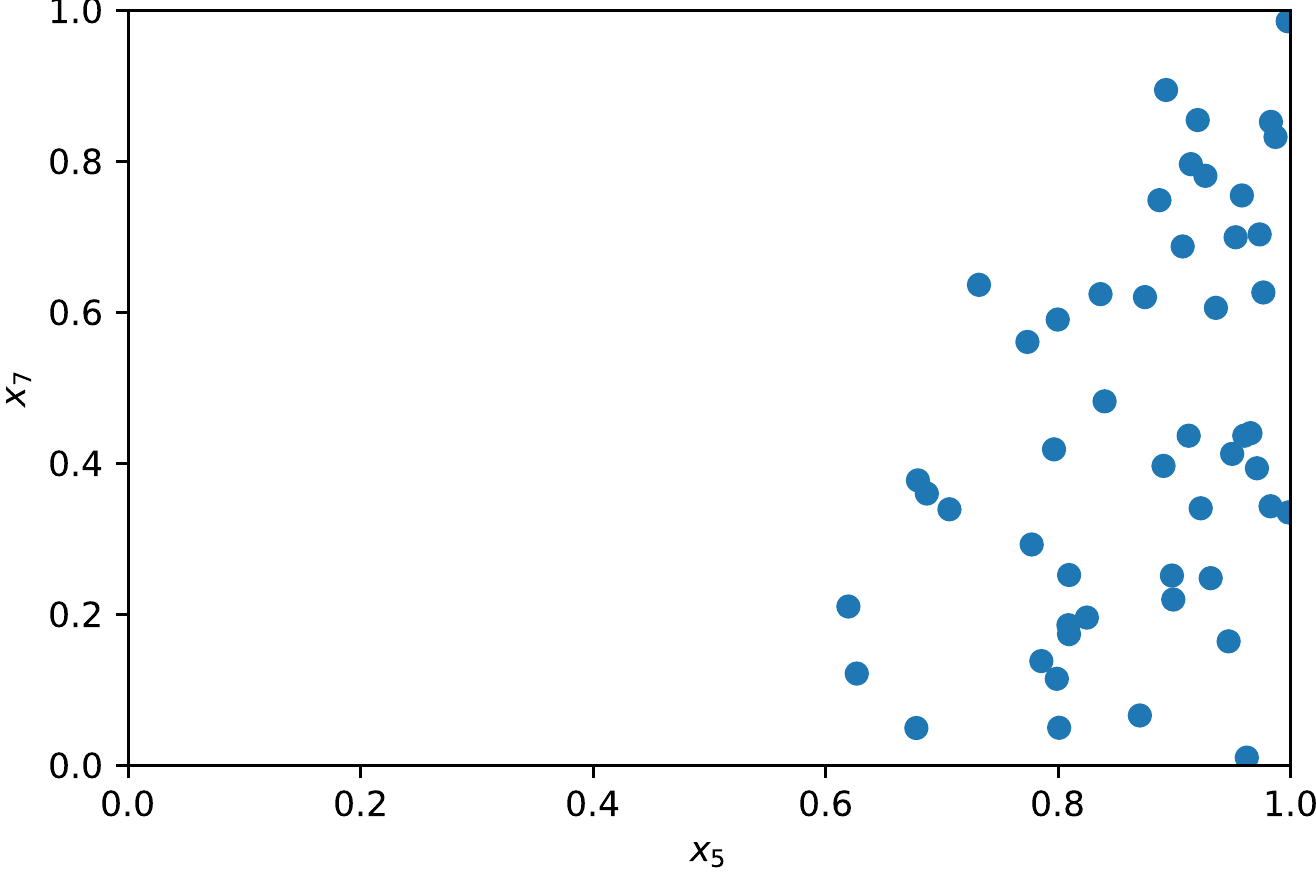}
  \caption{Example of the basin of attraction of the global minimum (Trial 1). The left panel shows that the basin is close to the $0$'s corner along the dimension of $x_1$ ($\text{mean} = 0.32, \text{SD} = 0.23$) and the dimension of $x_3$ ($\text{mean} = 0.32, \text{SD} = 0.21$). The right panel shows that the basin is close to the $1$'s corner along the dimension of $x_5$ ($\text{mean} = 0.87, \text{SD}  = 0.10$), but does not have any deviation along the dimension of $x_7$ ($\text{mean} = 0.44, \text{SD}  = 0.26$).}
  \label{Fig.6}
\end{figure}

\noindent First, the $t$-test is applied to the value of each variable (dimension) of those initial positions within the global minimum's basin of attraction in each trial out of 200 trials. The null hypothesis for the $t$-test is that those values are not different from $0.5$. In other words, we want to check across all dimensions to see if the global minimum's basin is shifted to a corner of the 0-1 hypercube. For example, if the projection of the global minimum's basin to the dimension of $x_1$ is near to $0$, it means that the global minimum's basin is close to the $0$'s corner of the 0-1 hypercube, no matter what the values are in other dimensions (see Figure \ref{Fig.6}).

We count for each dimension across 200 trials to see in how many trials the global minimum's basin of attraction deviates from $0.5$ in that dimension. The result (Table \ref{Tab.2}) evinces that this kind of deviation is widespread. This is a very important point, because the deviation of the global minimum's basin implies that the energy surface determined by the particular $C$ and $\boldsymbol{d}$ contains stable information for that dimension independent of the other dimensions. Hence, one can just determine some of the variables one at a time and not need to go back to adjust them again when processing other dimensions. This is totally different from the exhaustive search method since flipping one variable of the $\boldsymbol{x}$ would affect the value of $C \boldsymbol{x}$ and this would then impact the values of other dimensions. Therefore, one has to adjust the values of other dimensions again to try to obtain the correct object vector $\boldsymbol{d}$.
\begin{table}[htbp]
  \label{Tab.2}
  \centering
  \caption{Deviation of the basin of attraction of the global minimum. Each element in this table represents in how many trials out of 200 in total, the global minimum's basin deviated from $0.5$ in the corresponding dimension.}
  \begin{tabular}{cccccccccc}
  \toprule
  $x_1$ & $x_2$ & $x_3$ & $x_4$ & $x_5$ & $x_6$ & $x_7$ & $x_8$ & $x_9$ & $x_{10}$ \\
  \midrule
  153 & 143 & 147 & 157 & 160 & 145 & 151 & 158 & 149 & 160 \\
  \bottomrule
 \end{tabular}
\end{table}

For testing the correlations between each pair of two dimensions, we calculate the correlation matrix for those initial points within the global minimum's basin in each trial. After that, the number of pairs whose correlations are greater than $0.75$ or less than $-0.75$ are counted in each trial (the diagonal elements are excluded since all of them are $1$). The result indicates that only 17 trials out of 200 in total contain related pairs (two variables/dimensions are correlated with each other). This further suggests the independence of the global minimum's basin in different dimensions.

Nevertheless, it is quite hard to find a relationship between the augmented matrix $(C, \boldsymbol{d})$ and the location of the basin of attraction of the global minimum. Although the larger basin does exist, at this time we are unable to determine where it is, based on the problem itself. The relationship might or might not exist, but it needs to be studied further.

\section{Discussion}
\noindent The primary objective of this paper is to find a better continuous-time method to solve the 0-1 ILP feasibility problem without being trapped in a local well. For this purpose we conducted four experiments (Sections \ref{Sec.4}, \ref{Sec.5}, \ref{Sec.6}, \ref{Sec.7}) to analyze the solution procedure. When this continuous-time method is applied to a practical problem, the procedure should be as follows. (a) Section \ref{Sec.7} shows that the global minimum's basin of attraction is independent in some of the dimensions. If a future study could find a relationship between the augmented matrix $(C, \boldsymbol{d})$ and the location of the basin of the global minimum, one could just start by initializing a point for certain dimensions successively and perhaps not need to go back to adjust these basically well-determined dimensions. (b) Section \ref{Sec.6} shows that the global minimum attracting area is larger than $2^{-N}$, the discrete probability corresponding to exhaustive search. (c) Section \ref{Sec.4} suggests that our impulse algorithm is able to escape from the local minima to find the global minimum effectively and that its performance is better than escape by randomization. Thus, after randomly initializing a point, adopt the impulse algorithm to find the global minimum. (d) In Section \ref{Sec.5}, the time-to-solution distribution of the impulse algorithm shows us that more than half of the solutions could be solved in a short time. In other words, in the average case, the impulse algorithm is more likely to reach the solution compared with the exhaustive search method (the peak of the time-to-solution distribution skews to the left-hand side).

Another point that must be observed here is that, generally speaking, the time complexity of the impulse algorithm is still exponential. Although Ben-Hur proved the continuous-time ODE network has polynomial complexity \cite{ben2002theory}, our impulse algorithm includes the escape process, which makes the algorithm continue searching until it lands at the global minimum. Thus, the complexity of this algorithm is still exponential.

\section{Conclusion}
\noindent This paper proposed a continuous-time dynamical system for solving the 0-1 integer linear programming feasibility problem. First, we transformed the problem to a better form and proposed the impulse algorithm to tackle the local minima. Then, we discussed the time-to-solution distribution. After that, the basin of attraction of the global minimum and its location were investigated. However, if one wants to use this continuous-time method to solve problems, the relationship between the location of the basin of attraction and its corresponding problem  still need to be found, so this should be a direction of future work.
Our results here are empirical and preliminary. Future work will investigate simplification and other improvements to the impulse algorithm, relevant theorems, continuous-time complexity analysis, and exploration of performance on a wider range of instances.

\section*{Acknowledgement}
\noindent This project was completed during Chengrui Li's exchange period: An international student exchange program between Sichuan University and the University of Tennessee, Knoxville.

\bibliographystyle{unsrt}
\bibliography{ms}  

\end{document}